\documentclass{article}

\usepackage{PRIMEarxiv}

\usepackage[utf8]{inputenc} 
\usepackage[T1]{fontenc}    
\usepackage{hyperref}       
\usepackage{url}            
\usepackage{booktabs}       
\usepackage{amsfonts}       
\usepackage{nicefrac}       
\usepackage{microtype}      
\usepackage{lipsum}
\usepackage{fancyhdr}       
\usepackage{graphicx}       
\graphicspath{{media/}}     %

\usepackage{amsmath}
\usepackage{amsthm}
\usepackage{multirow}
\newtheorem{theorem}{Theorem}
\newtheorem{lemma}{Lemma}
\newtheorem{definition}{Definition}

\title{IC Mechanisms for Risk-Averse Advertisers in the Online Advertising System}

\author{
  Bingzhe Wang \\
  Gaoling School of Artificial Intelligence \\
  Renmin University of China \\
  Beijing, China \\
  \texttt{wbz2022@ruc.edu.cn} \\
   \And
  Ruohan Qian \\
  Baidu \\
  Beijing, China \\
  \texttt{qianruohan@baidu.com} \\
    \And
  Yuejia Dou \\
  Gaoling School of Artificial Intelligence \\
  Renmin University of China \\
  Beijing, China \\
  \texttt{douyuejia@ruc.edu.cn} \\
    \And
  Qi Qi \thanks{Corresponding author.} \\
  Gaoling School of Artificial Intelligence \\
  Renmin University of China \\
  Beijing, China \\
  \texttt{qi.qi@ruc.edu.cn} \\
    \And
  Bo Shen, Changyuan Li, Yixuan Zhang, Yixin Su, Xin Yuan, Wenqiang liu, Bin Zou \\
  Baidu \\
  Beijing, China \\
  \texttt{\{shenbo04,lichangyuan01,zhangyixuan05,suyixin,yuanxin06,liuwenqiang01,zoubin03\}@baidu.com} \\
    \And
  WEN YI, Zhi Guo, shuanglong li, LIU LIN  \\
  Baidu \\
  Beijing, China \\
  \texttt{\{yiwen01,guozhi,lishuanglong,liulin03\}@baidu.com} \\
}

\begin{document}
\maketitle

\begin{abstract}
The autobidding system generates huge revenue for advertising platforms, garnering substantial research attention. Existing studies in autobidding systems focus on designing \emph{Autobidding Incentive Compatible} (AIC) mechanisms, where the mechanism is \emph{Incentive Compatible} (IC) under ex ante expectations. However, upon deploying AIC mechanisms in advertising platforms, we observe a notable deviation between the actual auction outcomes and these expectations during runtime, particularly in the scene with few clicks (sparse-click). This discrepancy undermines truthful bidding among advertisers in AIC mechanisms, especially for risk-averse advertisers who are averse to outcomes that do not align with the expectations. To address this issue, we propose a mechanism, \emph{Decoupled First-Price Auction} (\textbf{DFP}), that retains its IC property even during runtime. \textbf{DFP} dynamically adjusts the payment based on real-time user conversion outcomes, ensuring that advertisers' realized utilities closely approximate their expected utilities during runtime. To realize the payment mechanism of \textbf{DFP}, we propose a PPO-based RL algorithm, with a meticulously crafted reward function. This algorithm dynamically adjusts the payment to fit \textbf{DFP} mechanism. We conduct extensive experiments leveraging real-world data to validate our findings.

\end{abstract}


\section{Introduction}
\label{intro}

According to statistics from the Interactive Advertising Bureau (IAB), online advertising has generated hundreds of billions of dollars in annual revenue for internet companies. In the traditional real-time bidding (RTB), each user query triggers an auction within the platform, where the auctioned items are the advertisement (ad) slots positioned from top to bottom on the search results page view (PV). Bidders in these auctions are advertisers interested in the specific query. During each auction, bidders independently submit their bids to the platform, considering various factors such as global constraints (e.g., budget, return on investment (ROI), or target cost per acquisition (tCPA)) and value per ad (e.g., per-click or per-conversion value). The platform then operates an auction mechanism to determine the allocation of slots and the payments made by the bidders based on their bids. Based on auction theory \cite{vickrey1961counterspeculation, clarke1971multipart, groves1973incentives, myerson1981optimal}, the earliest  mechanisms in RTB were the Generalized First Price (GFP) \cite{cary2007greedy, lahaie2007sponsored, edelman2007strategic, jansen2008sponsored} and Generalized Second Price (GSP) \cite{edelman2007internet} auctions.
With advancements in machine learning and the increasing availability of real-time data, an autobidding technology has gained growing attention and has been adopted in ad auctions \cite{aggarwal2019autobidding, babaioff2020non, deng2021towards, balseiro2021landscape, mehta2022auction}. In autobidding, bidders are value-maximizers, aiming to maximize the total value acquired over multiple rounds of auctions (rounds) under certain constraints. Bidders submit their constraints and values to the platform. The platform configures an autobidder for each bidder to bid on their behalf in each round.
Furthermore, the pricing model in the auction dictates when and how bidders pay. The earliest pricing model included
\emph{cost-per-click} (CPC) and \emph{cost-per-acquisition} (CPA), where advertisers bid/pay for
clicks and conversions (acquisitions), respectively.
Recently, \emph{optimized-cost-per-click} (OCPC), a model tailored to bidders' ultimate conversion goals,
has emerged. In OCPC, advertisers bid for conversions, while the platform charges them upon each click.
OCPC encapsulates the merits of both CPC and CPA, offering a mutually beneficial framework for both advertisers and platforms. By empowering advertisers to bid based on conversions, OCPC incentivizes platforms to deliver more precisely targeted advertisements, thereby enhancing conversion rates. Meanwhile, the click-based payment mechanism ensures that advertisers maintain greater control over their costs, while the platform secures a stable revenue stream.
Different from previous work which mainly focuses on CPC or CPA models, our work focuses on tCPA bidders in OCPC model.

Most research on autobidding has been focused on the ex ante properties of auctions, emphasizing bidders' utilities and constraints under expectation, with the objective of designing mechanisms that satisfy both \emph{Autobidding Incentive Compatibility} (AIC) and \emph{Individual Rationality} (IR) \cite{alimohammadi2023incentive, xing2023truthful}. Inspired by those related papers, we design an AIC mechanism, Coupled First-Price Auction (\textbf{CFP}), where the allocation mechanism accommodates the broadest spectrum of ranking scores, thereby ensuring flexibility and adaptability to cater to any allocation objectives of platforms.
 
\begin{figure}[t]
\centering
\includegraphics[width=0.5\textwidth]{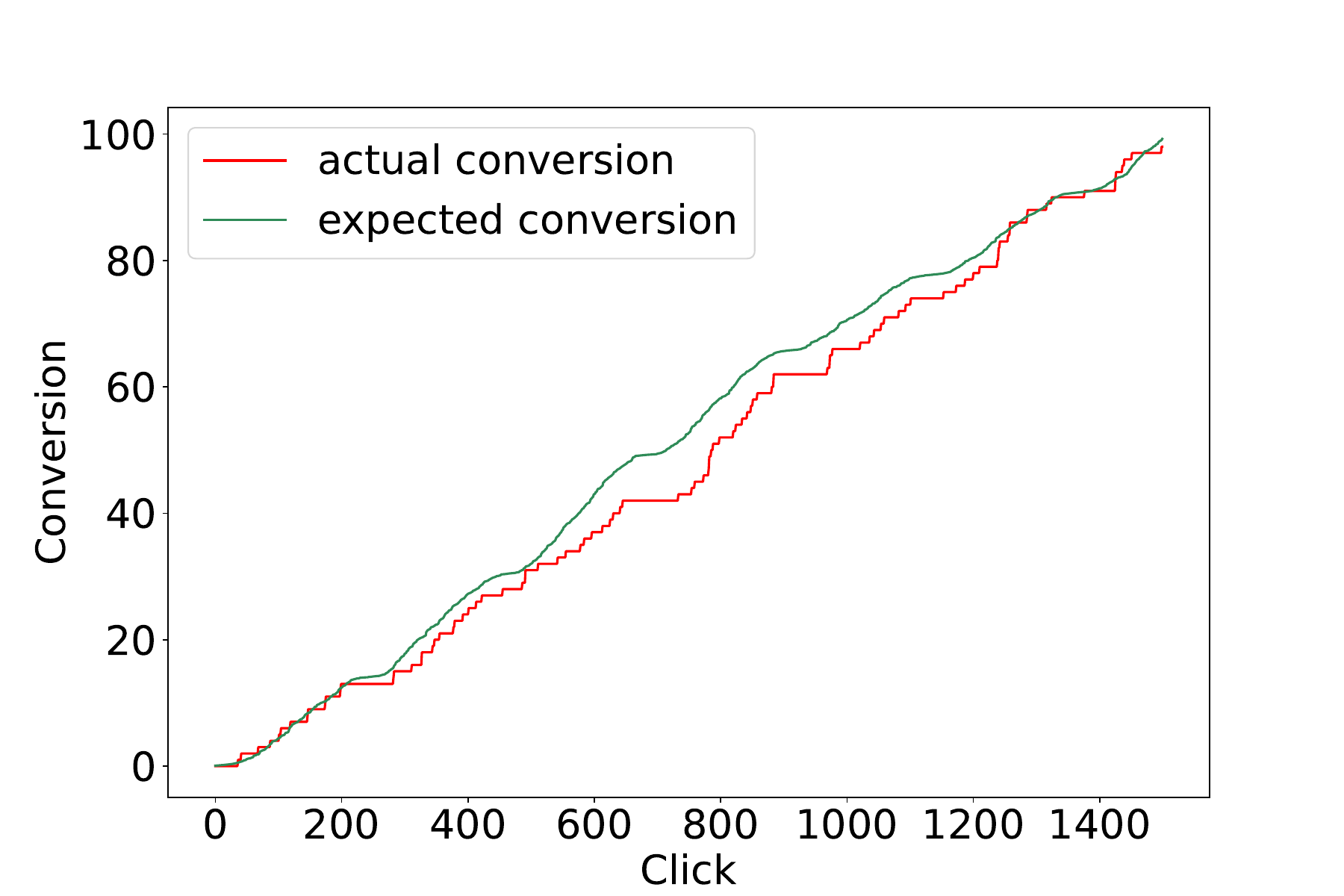}
\caption{The Conversion Volume of an Advertiser}
\label{demo}
\end{figure}

However, upon deploying \textbf{CFP}, we observed that bidders, particularly those sparse-click bidders, frequently resubmit bids, often with significant adjustments every day. Analyzing authentic data from our platform (exemplified in Figure \ref{demo} for a bidder, where the horizontal axis represents the change in click volume, and the vertical axis represents the conversion volume), we find that the bidder's actual conversion volumes closely align with expectations only when click volumes are substantial (after 1400 clicks). Conversely, in most instances  (before 1400 clicks), notable deviations between actual and expected conversions are observed. This divergence from expected conversions causes the utilities of these bidders in the AIC mechanism to deviate from its expectations.
For instance, as illustrated in Figure \ref{demo}, when the click volume reaches 800, the bidder observes that the actual conversion volume falls significantly short of her expectation. Consequently, her realized utility is less than anticipated, and the actual CPA surpasses her true tCPA. This discrepancy incentivizes her to submit a bid to the platform that is lower than her true tCPA.
Prior studies implicitly assumed that bidders are \emph{risk-neutral} (i.e., they only care about the expected outcomes without considering the variations during runtime). Specifically, they assume that bidders have no strategic behaviors that arise from the discrepancies between actual and expected outcomes across multiple auction rounds. However, our observation of bidders frequently resubmitting revised bids indicates that a number of bidders are \emph{risk-averse} (the most common risk preference type). A risk-averse bidder hopes her real-time utility to match the expectations. An outcome that does not meet expectations prompt risk-averse bidders to adjust their bids or even withdraw from the auction, leading to untruthful bidding and contributing to the instability of the AIC mechanism. We prove that risk-averse bidders will bid truthfully in an AIC mechanism only when the click volume is at least $\frac{(2+ \epsilon) \ln{\frac{1}{\epsilon}}}{\epsilon^2 \cdot \eta}$ for each bidder, where $\epsilon$ is a small positive constant, and $\eta$ denotes the maximum conversion rate (Lemma \ref{lemma5}). This requirement poses a challenge for both small and medium-sized bidders (clicks are sparse) and platforms. Put simply, the AIC mechanism works primarily for systems with sufficient click volumes, where real-time outcomes converge towards expectations.

\begin{figure}[ht]
\centering
\includegraphics[width=0.9\textwidth]{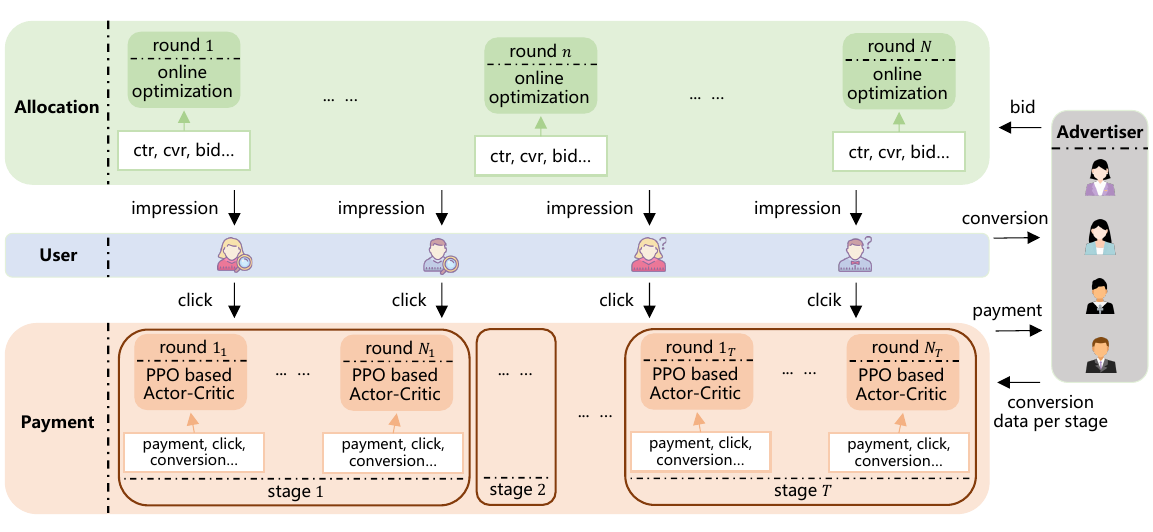}
\caption{Decoupled First-Price Auction}
\label{Decoupled Auction}
\end{figure}

To address the challenge of sparse clicks, we introduce a stricter notion of IC than AIC, named \emph{Time-Invariant Incentive Compatibility} (TIC), in which risk-averse bidders bid truthfully. We characterize the conditions for a mechanism to satisfy TIC (Lemma \ref{lemma6}), and desgin a TIC mechanism, Decoupled First-Price Auction (\textbf{DFP}, see Figure \ref{Decoupled Auction}).
Traditional mechanisms have a direct mapping from bids to per-click payments, such as GFP sets the payments equal to corresponding bids and GSP sets the payments equal to the next highest bid. Instead, in \textbf{DFP}, the bids in each round are not explicitly mapped to the per-click payments. Namely, the allocation and payment mechanisms operate independently.
In this paper, we demonstrate the theoretical feasibility of decoupling the intricate problem of mechanism design into two distinct, yet interdependent, online optimization problems, aiming to design allocation mechanisms and payment mechanisms, respectively. The allocation mechanism is to maximize certain objectives of the platform. We insert real-time outcomes from users and bidders (even delayed feedback is allowed) between the allocation mechanism and the payment mechanism, and adjust the per-click payments in the payment mechanism accordingly. This approach allows our mechanism not only rely on expectations, but also incorporate users' real-time outcomes into the payment mechanism to satisfy TIC (Theorem \ref{theorem2}).

The allocation mechanism of \textbf{DFP} is a canonical online optimization problem, which can be addressed utilizing various algorithms such as online algorithms, deep learning, and reinforcement learning (RL), among others \cite{mehta2007adwords, devanur2009adwords}. As we can leverage any allocation mechanism in DFP, it is not the focus of this paper and we focus on the payment mechanism design. Distinct from the majority of online decision-making problems that solely necessitate pricing decisions, the payment mechanism of \textbf{DFP} necessitates two task: firstly, leveraging historical data to forecast the actual conversion outcomes of future users, and secondly, adjusting the per-click payment accordingly based on these forecasts. To realize this intricate payment mechanism in \textbf{DFP}, we propose a PPO-based RL algorithm, with meticulously crafted reward function. 

Our contributions can be summarized as follows:

\textbf{Propose an AIC Mechanism.} We investigate the tCPA bidders in OCPC, demonstrating that mechanisms with first-price payment rule are AIC. Subsequently, we propose an AIC mechanism \textbf{CFP}, which supports any allocation objective of the platform.

\textbf{Introduce a New Concept.} We find bidders untruthful bidding because of the deviations of actual outcomes from their expectations in AIC mechanisms, particularly in sparse-click systems. This anomaly arises due to the ignorance of bidders' risk preferences. We introduce a new concept, termed \emph{Time-Invariant Incentive Compatibility} (TIC), which ensures that risk-averse bidders will engage in truthful bidding. 
To the best of our knowledge, we are the first to consider bidders' risk preferences and the real-time bidding processes in autobidding scenarios.

\textbf{Design a TIC Mechanism by Decoupling.} To address the challenge of sparse-click systems, we derive the conditions for TIC. By these conditions, we design \textbf{DFP}, a TIC mechanism whose allocation and payment mechanisms are designed by two distinct, yet interdependent, online optimization problems. Our decoupled approach showcases superior game-theoretic properties compared to mechanisms designed holistically, stemming from its inherent ability to integrate users' actual behaviors into the formulation of payment mechanisms.

\textbf{Implement a RL-based Payment Mechanism.} We employ a PPO-based RL to construct the payment mechanism. Through meticulous design of the reward function, our payment mechanism dynamically adjusts per-click payments based on historical bidding outcomes and the anticipating of future user conversions.

\textbf{Conduct Extensive Experiments.} Extensive experiments on vast real-world online data validate that \textbf{DFP} maintains robust IC properties even in sparse-click systems, with small payment fluctuation of bidders.

\section{Further Related Work}

Among the earliest research of autobidding, Aggarwal et al. introduced the optimal bidding strategy within truthful mechanisms \cite{aggarwal2019autobidding}, coined as uniform bidding, which operates as a pacing algorithm aimed at smoothly depleting budgets \cite{balseiro2019learning, conitzer2022pacing, chen2023complexity, conitzer2022multiplicative, gaitonde2022budget}. Analogously, shading algorithms have emerged where bidders adopt coefficients less than unity to obscure their genuine valuations \cite{hortaccsu2018bid, karlsson2021adaptive, gong2023mebs, zhang2021meow}. Beyond these optimization-theoretic bidding algorithms, data-driven approaches have also surfaced, further diversifying the autobidding landscape \cite{chen2023model, zhang2023personalized}.

Parallel to bidding strategies, mechanism design constitutes another pivotal research direction in autobidding. Golrezaei et al. devised revenue-maximizing auctions tailored for bidders subject to ROI constraints \cite{golrezaei2021auction}. Balseiro et al. then introduced robust mechanisms based on VCG and GSP with reserve prices \cite{balseiro2021robust}. Subsequently, they delved into the performance of bidders with varying objectives within these revenue-maximizing mechanisms \cite{balseiro2021landscape}. Building upon this foundation, Balseiro et al. demonstrated that for bidders constrained by ROI, the optimal mechanism remains consistent across bidders' targets, thereby obviating the need for personalized mechanisms tailored to specific objectives \cite{balseiro2023optimal}. Alimohammadi et al. proposed the notion of AIC to characterize whether a mechanism is IC in expectation \cite{alimohammadi2023incentive}. The work most pertinent to ours is that of Lv et al., who shifted the focus from a mechanism's ex-ante properties to examining bidders' ex post ROI constraints within a randomized mechanism \cite{lv2023auction}. However, their analysis presupposes that bidders' ROI is public information and bidders possess quasi-linear utilities. On the contrary, our study posits bidders as value-maximizers, reflecting the prevalent utility function in autobidding scenarios, and assumes that bidders' values are private, aligning with the practical intricacies of real-world markets.

\section{Model and Preliminary}
\label{model}

In the context of online advertising, every time a user queries the platform, an auction is orchestrated among $M$ bidders for the opportunity to show their ads. $K$ winning bidders are displayed from top to bottom in the PV presented to the user. Over a discrete time span of $N$ rounds, we denote the sets of bidders, rounds, and slots by $[M]$, $[N]$, and $[K]$ respectively, with $m$, $n$, and $k$ representing a specific bidder, round, and slot.

An auction mechanism $\mathcal{M}(\boldsymbol{x}, \boldsymbol{p})$ comprises an allocation rule $\boldsymbol{x}: \mathbb{R}^{M} \rightarrow [0, 1]^{M \times N \times K}$ and a payment rule $\boldsymbol{p}: \mathbb{R}^{M} \rightarrow \mathbb{R}^{M \times N \times K}$. Specifically, $x_{mnk}$ indicates the allocation outcome, where $x_{mnk} = 1$ if bidder $m$ is placed on slot $k$ in round $n$, and $x_{mnk} = 0$ otherwise. The user clicks the displayed bidders with a certain probability, which subsequently lead the user to leave the PV and enter the bidder's landing page, potentially resulting in a purchase, termed acquisition or conversion. The click-through rate (CTR) and conversion rate (CVR) of bidder $m$ on slot $k$ in round $n$ are denoted as $ctr_{mnk}$ and $cvr_{mnk}$, respectively. Binary variables $y_{mnk}$ and $z_{mnk}$ capture click and conversion outcomes, where $y_{mnk} = 1$ signifies a click and $y_{mnk} = 0$ otherwise. $z_{mnk} = 1$ only if a conversion occurs, accompanied by a value $v_{mnk}$ to bidder $m$; otherwise, $z_{mnk} = 0$. In ad systems, it is typically that $1 \geq ctr_{mnk} \gg cvr_{mnk} \geq 0$, and $ctr_{mn1} \geq ctr_{mn2} \geq \cdots \geq ctr_{mnK}, \forall m, \forall n$. In line with autobidding literature, both $ctr_{mnk}$ and $cvr_{mnk}$ are considered common knowledge shared between the platform and bidder $m$, and these rates are precisely estimated.

The impression, click, and conversion vloume for bidder $m$ up to round $n$ are represented by $X_{mn} = \sum_{j \in [n], k \in [K]} x_{mjk}$, $Y_{mn} = \sum_{j \in [n], k \in [K]} y_{mjk}$, and $Z_{mn} = \sum_{j \in [n], k \in [K]} z_{mjk}$, respectively. The payment of bidder $m$ for slot $k$ in round $n$ is denoted as $p_{mnk}$, and the total payment across $n$ rounds is $P_{mn} = \sum_{j \in [n], k \in [K]} p_{mjk}$. Consequently, the CPA of bidder $m$ over $n$ rounds is $CPA_{mn} = \frac{P_{mn}}{Z_{mn}}$. Bidder $m$ has a tCPA denoted by $tCPA_{m}$, which represents the average amount the bidder willing to pay for a conversion. Both $v_{mnk}$ and $tCPA_{m}$ are private information of bidder $m$.

Specifically, bidder $m$ submits a bid $b_{m}$ to the platform, claiming that $b_{m}$ equates to $tCPA_{m}$. However, $b_{m}$ may not necessarily equal $tCPA_{m}$ due to bidders' strategic motivations to achieve higher utility through untruthful bidding. We define $\boldsymbol{b} = (b_{1}, \ldots, b_{m}, \ldots, b_{M})$ as the bidding profile of all ads, and $\boldsymbol{b}_{-m} = (b_{1}, \ldots, b_{m-1}, b_{m+1}, \ldots, b_{M})$ is the bidding profile excluding bidder $m$. We focus on value-maximizers, a prevalent bidder type encountered both theoretically and practically within autobidding frameworks. Value-maximizers submit their constraints, such as budget, ROI, and tCPA, as bids, with the objective of maximizing the total value acquired over multiple rounds.

\begin{definition}[Value-Maximizer]\label{definition2}

The utility $u_{mn}$ of a value-maximizer $m$ is the aggregate conversion value derived from $n$ auction rounds: $u_{mn}(b_{m}, \boldsymbol{b}_{-m}) = \sum_{j \in [n], k \in [K]}{v_{mjk} \cdot z_{mjk}}$.

\end{definition}

Consequently, if bidders are value-maximizers, their objective is to maximize conversion value while ensuring that the CPA does not exceed the tCPA. To this end, each bidder $m$ strategically submits the optimal solution to the optimization problem $\boldsymbol{P}_{m}$ as their bid, thereby maximizing their utility.

\begin{align*}\tag{$\boldsymbol{P}_{m}$}
\max \quad u_{mN}(b_{m}, \boldsymbol{b}_{-m}) \\
\text{s.t.} \quad CPA_{mN} = \frac{P_{mN}}{Z_{mN}} & \le tCPA_{m}, && \\
\sum_{k \in [K]}{x_{mnk}} & \le 1, && \forall n \in [N] \\
x_{mnk} & \in \{0, 1\}, && \forall n \in [N], k \in [K].
\end{align*}

In the mechanism design theory, a mechanism $\mathcal{M}$ that satisfies both \emph{incentive compatibility} (IC) and \emph{individual rationality} (IR) ensures that rational bidders' optimal strategy is to truthfully reveal their private information. In autobidding, the notion of \emph{Autobidding Incentive Compatibility} serves as a pertinent metric analogous to IC in this specific setting, assessing whether value-maximizers engage in truthful bidding, i.e., $b_{m} = tCPA_{m}$. This framework ensures bidders have no incentive to deviate from truthful bidding, thereby promoting stability and efficiency in autobidding. Throughout this paper, we employ the notation $\overline{e}$ and $\widehat{e}$ to signify the expected value and actual value of the random variable $e$.

\begin{definition}[Autobidding Incentive Compatibility (AIC)\cite{alimohammadi2023incentive}]\label{definition3}

A mechanism $\mathcal{M}(\boldsymbol{x}, \boldsymbol{p})$ is said to be Autobidding Incentive Compatibility iff for every bidder $m$ and for any bidding profile $\boldsymbol{b}_{-m}$, the bidder's expected utility when bidding truthfully ($b_{m} = tCPA_{m}$) is no less than her expected utility under any alternative bid $b_{m}' \in \mathbb{R}$:

$$
\overline{u}_{mN}(tCPA_{m}, \boldsymbol{b}_{-m}) \ge \overline{u}_{mN}(b_{m}', \boldsymbol{b}_{-m}), \quad \forall b_{m}' \in \mathbb{R}.
$$

\end{definition}

\begin{definition}[Individual Rationality (IR)]\label{definition4}

A mechanism $\mathcal{M}(\boldsymbol{x}, \boldsymbol{p})$ is said to be Individual Rationality iff for every bidder $m$, her expected CPA ($\overline{CPA}_{mN}$) does not exceed $tCPA_{m}$: $\overline{CPA}_{mN} \le tCPA_{m}$.

\end{definition}

These definitions form the cornerstone of our analysis, ensuring that the proposed autobidding mechanisms not only incentivize truthful bidding but also uphold the fundamental economic principles of rationality and efficiency.

\section{IC Mechanism Design}
\label{tic}

In this section, we introduce an AIC mechanism, \textbf{CFP}, and a TIC mechanism, \textbf{DFP}.

\subsection{Coupled First-Price Auction}
\label{cfp}

In this subsection, we introduce \textbf{CFP}, an AIC mechanism whose allocation rule accommodates any form of ranking score $r_{mnk}$. This flexibility allows for definitions akin to GFP snd GSP, where $r_{mnk} = b_{m} \cdot ctr_{mnk}$. Alternatively, the ranking score can be implicitly represented as the optimal solution to the platform's allocation objective, for instance, when the platform aims to maximize social welfare (SW), wherein $r_{mnk}$ can be defined as the solution to the Adwords problem \cite{mehta2007adwords}. Consequently, as long as $r_{mnk}$ is a weakly increasing function of $b_m$, \textbf{CFP} satisfies arbitrary allocation objectives, extending even to multi-objective settings. Furthermore, we establish that within OCPC model with tCPA bidders, any AIC mechanism adopts a first-price payment rule. Lastly, we derive the equivalent conditions for AIC, which serve as a foundation for our subsequent design of TIC mechanisms.

\begin{definition}[Coupled First-Price Auction (\textbf{CFP})]\label{definition1}

The allocation and payment mechanism of the Coupled First-Price Auction are defined as follows. Allocation Mechanism $\boldsymbol{x}$:

$$
x_{mnk}=
\begin{cases}
1, & \text{if } r_{mnk} = r_{mn(k)}, \\
0, & \text{otherwise},
\end{cases}
$$
where $r_{mn(k)}$ is the $k$-th highest ranking score within vector $\boldsymbol{r}_n$.

Payment Mechanism $\boldsymbol{p}$: $p_{mnk} = b_{m} \cdot y_{mnk} \cdot cvr_{mnk}$.

\end{definition}

To analyze \textbf{CFP}, we first express the expected form of the random variable in Section \ref{model} according to the definition:

$$
\overline{y}_{mnk} = x_{mnk} \cdot ctr_{mnk}, \quad \overline{Y}_{mn} = \sum_{j \in [n], k \in [K]}{\overline{y}_{mjk}},
$$

$$
\overline{z}_{mnk} = \overline{y}_{mnk} \cdot cvr_{mnk}, \quad \overline{Z}_{mn} = \sum_{j \in [n], k \in [K]}{\overline{z}_{mjk}},
$$

$$
\overline{p}_{mnk} = b_{m} \cdot \overline{y}_{mnk} \cdot cvr_{mnk}, \quad \overline{P}_{mn} = \sum_{j \in [n], k \in [K]}{\overline{p}_{mjk}},
$$

$$
\overline{u}_{mn}(b_{m}, \boldsymbol{b}_{-m}) = \sum_{j \in [n], k \in [K]}{v_{mjk} \cdot \overline{z}_{mjk}}, \quad \overline{CPA}_{mn} = \frac{\overline{P}_{mn}}{\overline{Z}_{mn}}.
$$

Before the $N$-round auction, bidder $m$ determines her optimal bidding strategy by solving the expected problem $\overline{\boldsymbol{P}}_{m}$ corresponding to $\boldsymbol{P}_{m}$.

\begin{align*}\tag{$\overline{\boldsymbol{P}}_{m}$}
\max \quad \overline{u}_{mN} & (b_{m}, \boldsymbol{b}_{-m}) \\
\text{s.t.} \quad \overline{CPA}_{mN} & \le tCPA_{m}, && \\
\sum_{k \in [K]} x_{mnk} & \leq 1, && \forall n \in [N] \\
x_{mnk} & \in \{0, 1\}, && \forall n \in [N], k \in [K].
\end{align*}

\begin{lemma}\label{lemma1}

In \textbf{CFP}, the optimal bidding strategy for tCPA bidder $m$ is $b_{m}^{*} = tCPA_{m}$. Furthermore, if all bidders are tCPA bidders, then the strategy $b_{m}^{*} = tCPA_{m}$ constitutes a Nash Equilibrium (NE).

\end{lemma}

As truthful bidding is the optimal strategy for all bidders, i.e. $\forall m, b_{m}^{*} = tCPA_{m}$ is a NE, therefore \textbf{CFP} is AIC.

\begin{theorem}\label{theorem1}

In OCPC with tCPA bidders, \textbf{CFP} is AIC and IR.

\end{theorem}

\begin{lemma}\label{lemma2}

In OCPC model with tCPA bidders, \textbf{CFP} is both AIC and IR iff $\overline{CPA}_{mN} = tCPA_{m}$.

\end{lemma}

\subsection{Decoupled First-Price Auction}
\label{dfp}

In Section \ref{cfp}, we established that truthful bidding is the optimal strategy for bidders to maximize their expected utility before engaging the auction. Nevertheless, we find that the real-time outcomes during runtime, and even after $N$ rounds of the auction, can significantly deviate from these expectations, particularly in sparse-click systems. Prior research implicitly assumed bidders to be risk-neutral, failing to strategically adapt their behaviors in response to observed deviations between actual outcomes and expectations across multiple auction rounds. This underlying assumption contradicts realistic ad systems, wherein the auction process is a common knowledge shared between platforms and bidders. Consequently, a more practical problem formulation necessitates considering bidders' risk preferences and the strategic behaviors that align with these preferences. Our work stands as the first to incorporate bidders' risk preferences into the analysis of auction dynamics, adopting risk-aversion, the most prevalent risk preference, as the preference for bidder behavior. Risk-averse bidders periodically evaluate the auction mechanism's performance over multiple rounds, requiring that the mechanism's actual outcomes align with their ex ante expectations. Failure to meet this criterion may prompt bidders to adjust their bids or even withdraw from the auction altogether, potentially destabilizing the ad system.

By definition, $\widehat{Y}_{mn} = \sum_{j \in [n], k \in [K]}{\widehat{y}_{mjk}},$

$$
\widehat{Z}_{mn} = \sum_{j \in [n], k \in [K]}{\widehat{z}_{mjk}}, \quad \widehat{P}_{mn} = \sum_{j \in [n], k \in [K]}{\widehat{y}_{mjk} \cdot \widehat{p}_{mjk}},
$$

$$
\widehat{u}_{mn} = \sum_{j \in [n], k \in [K]}{v_{mjk} \cdot \widehat{z}_{mjk}}, \quad \widehat{CPA}_{mn} = \frac{\widehat{P}_{mn}}{\widehat{Z}_{mn}}.
$$

\begin{definition}[Risk-Averse Bidder]\label{definition5}

An bidder is a risk-averse bidder if she adjusts her bids or withdraws from the auction when she observes deviations between actual and expected outcomes during the auction process.

\end{definition}

The risk-averse bidders partition the $N$ rounds into $T$ stages, and observes the real-time outcomes of the mechanism after each stage. Let $n_t$ denote the $n$-th round in stage $t$, and $N_t$ is the amount of rounds in stage $t$, with $\sum_{t \in [T]} N_t = N$. Based on Lemma \ref{lemma2}, we derive the condition for a risk-averse bidder to maintain truthful bidding:

\begin{lemma}\label{lemma3}

In the mechanism $\mathcal{M}$, a risk-averse bidder $m$ will bid truthfully during the auction process iff $\frac{tCPA_m}{\widehat{CPA}_{mN_t}} = 1, \forall t \in [T]$.

\end{lemma}

Analogously, we introduce the concept of \emph{$\epsilon$-risk-averse bidder}: a $\epsilon$-risk-averse bidder will bid truthfully during the auction process iff $1 - \epsilon \leq \frac{tCPA_m}{\widehat{CPA}_{mN_t}} \leq 1 + \epsilon, \forall t \in [T]$. In this paper, $\epsilon$ is a small non-negative constant.

Then, we introduce the concept of TIC to evaluate whether a mechanism maintains IC during the auction process.

\begin{definition}[Time-Invariant Incentive Compatibility (TIC)]\label{definition6}

A mechanism $\mathcal{M}$ is Time-Invariant Incentive Compatibility iff truthful bidding is the optimal strategy for all risk-averse bidders during the auction process.

\end{definition}

Based on Lemma \ref{lemma3} and Definition \ref{definition6}, we have derived the properties of TIC mechanisms.

\begin{lemma}\label{lemma4}

In OCPC model with tCPA bidders, a mechanism $\mathcal{M}$ is TIC iff $\frac{tCPA_m}{\widehat{CPA}_{mN_t}} = 1, \forall m \in [M],  \forall t \in [T]$.

\end{lemma}

Analogously, we introduce the notion of \emph{$\epsilon$-TIC}: a mechanism $\mathcal{M}$ is $\epsilon$-TIC iff truthful bidding is the optimal bidding strategy for all $\epsilon$-risk-averse bidders during the auction process.

It is evident that the TIC criterion constitutes a stricter definition than AIC: every TIC mechanism is inherently AIC, but the converse is not necessarily true. Lemma \ref{lemma5} elucidates the conditions under which an AIC mechanism can satisfy $\epsilon$-TIC, emphasizing the necessity of substantial click volumes for each bidder across all stages.

\begin{lemma}\label{lemma5}

To ensure that a mechanism $\mathcal{M}(\boldsymbol{x}, \boldsymbol{p})$ satisfies $\epsilon$-TIC with a probability of at least $1 - \epsilon$, it is necessary that $\widehat{Y}_{mN_{t}} \ge \frac{(2+ \epsilon) \ln{\frac{1}{\epsilon}}}{\epsilon^2 \cdot \eta}, \forall \epsilon > 0$, where $\eta = \max_{m \in [M], n \in [N], k \in [K]} \{ cvr_{mnk} \}$.

\end{lemma}

It is noteworthy that the CPA model with tCPA bidders inherently exhibits TIC properties since bidder $m$ are only charged $tCPA_{m}$ for clicks when $ y_{mnk} = z_{mnk} = 1$, and no fee is incurred for clicks without conversions ($y_{mnk} = 1, z_{mnk} = 0$). However, besides the advantages of OCPC over CPA outlined in Section \ref{intro}, CPA remains ineffective in addressing the issue of feedback delay: platforms must charge bidders immediately upon clicks, whereas conversions, occurring significantly later and off-platform, require bidders to relay conversion outcomes back to the platform. This delay poses challenges for CPA execution. In contrast, OCPC offers greater control over the billing process, optimizing the bidder's billing experience. Lemma \ref{lemma5} establishes a lower bound on the required of click volume for the AIC mechanism to satisfy $\epsilon$-TIC. As $\epsilon$ decreases, the probability of the mechanism satisfying $\epsilon$-TIC increases, thereby bringing $\epsilon$-TIC closer to the ideal TIC. However, this reduction in $\epsilon$ necessitates an significant increase in the lower bound of the click volume. This poses a challenge for sparse-click systems, as the daily click volume of most bidders often do not meet these thresholds in real-world scenarios. Consequently, it is necessary to modify \textbf{CFP} to minimize the discrepancy between bidders' CPA and tCPA for designing TIC mechanisms.

\begin{lemma}\label{lemma6}

In OCPC model with tCPA bidders, \textbf{CFP} is TIC iff $\widehat{Z}_{mN_t} \cdot tCPA_{m} = \sum_{n \in [N_t], k \in [K]}{\widehat{y}_{mnk} \cdot \widehat{p}_{mnk}}, \forall m, t$.

\end{lemma}

By Lemma \ref{lemma6}, to ensure that \textbf{CFP} satisfies TIC, the platform confronts an online optimization problem $\widehat{\boldsymbol{P}}_{Plt}$. In this problem, the platform must make irrevocable decisions, $\widehat{p}_{mnk}$, instantaneously upon each user click. Based on this, we devise \textbf{DFP}.

\begin{align*}\tag{$\widehat{\boldsymbol{P}}_{Plt}$}
\min & \quad \sum_{m \in [M]}{\left| \frac{\widehat{Z}_{mN_t} \cdot tCPA_{m}}{\sum_{n \in [N_t], k \in [K]}{\widehat{y}_{mnk} \cdot \widehat{p}_{mnk}}} - 1 \right|} \\
\text{s.t.} & \quad \widehat{p}_{mnk} \ge 0, \quad \forall m \in [M], n \in [N_t], k \in [K].
\end{align*}

\begin{definition}[Decoupled First-Price Auction (\textbf{DFP})]

The allocation and payment mechanism of the Decoupled First-Price Auction are defined as follows. Allocation Mechanism $\boldsymbol{x}$:

$$
x_{mnk}=
\begin{cases}
1, & \text{if } r_{mnk} = r_{mn(k)}, \\
0, & \text{otherwise},
\end{cases}
$$
where $r_{mn(k)}$ is the $k$-th highest ranking score within vector $\boldsymbol{r}_n$.

Payment Mechanism $\boldsymbol{p}$: $\widehat{p}_{mnk} = p_{mnk}^{*}$, where $p_{mnk}^{*}$ is the optimal solution to $\widehat{\boldsymbol{P}}_{Plt}$.

\end{definition}

By $\widehat{p}_{mnk} = p_{mnk}^{*}$, we have $\sum_{m \in [M]}{\left| \frac{\widehat{Z}_{mN_t} \cdot tCPA_{m}}{\sum_{n \in [N_t], k \in [K]}{\widehat{y}_{mnk} \cdot \widehat{p}_{mnk}}} - 1 \right|} = 0$. Then, \textbf{DFP} is TIC.

\begin{theorem}\label{theorem2}

In OCPC model with tCPA bidders, \textbf{DFP} is TIC.

\end{theorem}

In \textbf{DFP}, the platform can formulate online allocation problems based on its allocation objectives, such as the Adwords problem aimed at maximizing social welfare \cite{mehta2007adwords}, and subsequently solve this online problem to derive corresponding ranking scores $\boldsymbol{r}_n$. Many algorithms exist for solving online allocation problems, capable of achieving near-optimal allocation \cite{devanur2009adwords}, thereby suggesting that both \textbf{CFP} and \textbf{DFP} can attain optimal allocation efficiency. The actual conversion outcomes of users decouple the allocation and payment mechanisms in \textbf{DFP}, making an implicit mapping between per-click payments and bids, dynamically adjusting based on bidders' feedback rather than an explicit mapping. However, due to the inherent delay in conversion data feedback relative to the platform's payment decisions, the platform lacks knowledge of $\widehat{Z}_{mN_t}$ when determining $\widehat{p}_{mnk}$. Consequently, the platform must concurrently estimate $\widehat{Z}_{mN_t}$ and solve for $p_{mnk}^{*}$, posing a formidable challenge.

Assuming knowledge of $\widehat{Z}_{mN_t}$, $\widehat{\boldsymbol{P}}_{Plt}$ reduces to a standard online decision-making problem, which can be tackled by directly using online algorithms, deep learning, or RL. Nevertheless, within \textbf{DFP}, the payment mechanism necessitates the estimation of $\widehat{Z}_{mN_t}$. To mitigate the impact of inaccuracies of $\widehat{Z}_{mN_t}$ on \textbf{DFP}, we meticulously define the reward function, devising a PPO-based Actor-Critic algorithm. In next section, we elaborate on how RL can be leveraged to learn the payment mechanism in \textbf{DFP}.

\section{Payment Mechanism Design}
\label{payment}

In this section, we demonstrate the design of the RL-based payment mechanism of \textbf{DFP}.

\subsection{Markov Decision Processes}

Given the formidable capabilities of RL in handling high-dimensional data, we formulate the problem of payment mechanism design of \textbf{DFP} as a Markov Decision Process (MDP) and employ RL to adjust per-click payment. The MDP is represented by the tuple $<\mathcal{S}, \mathcal{A}, \mathcal{R}, \gamma>$, where $\mathcal{S}$, $\mathcal{A}$, and $\mathcal{R}$ denote the space of states, actions (i.e., payments), and rewards, respectively, and $\gamma \in [0, 1]$ is the discount factor. We use a model-free RL to tackle this problem, thereby obviating the need for an explicit model of the transition dynamics.

$\mathcal{S}$: The state $\boldsymbol{s}_{n} = (s_{1n}, \cdots, s_{Mn}) \in \mathcal{S}$ at round $n$ encapsulates a comprehensive array of expected and actual click, conversion, and payment information for all bidders. Specifically, $s_{mn}$ comprises elements such as $\widehat{y}_{mnk}$, $\widehat{Y}_{mn}$, $\overline{Z}_{mn}$, $\widehat{Z}_{mn}$, $\overline{P}_{mn}$, $\widehat{P}_{mn}$, $tCPA_{m}$, $\sum_{j \in [n], k \in [K]}{\widehat{y}_{mjk} \cdot \widehat{p}_{mjk}}$, and $\widehat{p}_{mn_{m}k}$, where $\widehat{p}_{mn_{m}k}$ signifies the last non-zero payment of bidder $m$, corresponding to the payment for the last click (as payments are zero for non-click impressions). Notably, $\widehat{Z}_{mn}$ does not update in real-time with $n$ but is updated $T$ times throughout $N$ rounds, once at the end of each stage $t$, implying $\widehat{Z}_{m1_{t}} = \widehat{Z}_{m2_{t}} = \cdots, \widehat{Z}_{mN_{t}}, \forall t$.

$\mathcal{A}$: At round $n$, based on the current state $\boldsymbol{s}_{n}$, the algorithm generates an action $\widehat{\boldsymbol{p}}_{n} = (\widehat{p}_{1n}, \cdots, \widehat{p}_{Mn}) \in \mathcal{A}$, where $\widehat{p}_{mn} = \widehat{p}_{mnk} \in \mathbb{R}^{+}, \forall k$, represents the payment of bidder $m$ in round $n$. Importantly, $\widehat{p}_{mn}$ is non-zero only for clicked bidders.

$\mathcal{R}$: Upon executing action $\widehat{\boldsymbol{p}}_{n}$ in state $\boldsymbol{s}_{n}$, the algorithm receives a reward $r_{n}: \mathcal{S} \times \mathcal{A} \to \mathcal{R} \in \mathbb{R}$. As elucidated in Section \ref{dfp}, to ensure \textbf{DFP} is TIC, the algorithm must concurrently learn action $\widehat{p}_{mn}$ and estimate $\widehat{Z}_{mN_t}$, posing a distinct challenge from prior problems that solely focused on learning $\widehat{p}_{mn}$. To address this complexity, we define the reward $r_{n}$ into two components: $r_{n} = r_{n}^{(1)} + \zeta r_{n}^{(2)}$, where the hyperparameter $\zeta$ controls the relative weights of these two objectives. Specifically, $r_{n}^{(1)}$ incentivizes the algorithm to converge towards a more accurate optimal solution for $\widehat{\boldsymbol{P}}_{Plt}$, reinforcing the precision of the learned payment. Meanwhile, $r_{n}^{(2)}$ serves to reduce fluctuations in payments, mitigating the adverse effects of erroneous conversion estimation, and fostering a more stable payment experience for bidders. In essence, $r_{n}^{(2)}$ mitigates the risks associated with imprecise conversion forecasts by rewarding the algorithm for achieving a more precise $\widehat{Z}_{mN_t}$. The meticulously designed reward function facilitates the algorithm's TIC.

$$
r_{n}^{(1)} = -\log{\sum_{m \in [M]}{| \frac{\sum_{n \in [N_t], k \in [K]}{\widehat{y}_{mnk} \cdot \widehat{p}_{mnk}}}{\widehat{Z}_{mN_t} \cdot tCPA_{m} + \xi} - 1|}},
$$
where, $\xi$ is a constant introduced to prevent the denominator from being zero.

$$
r_{n}^{(2)} = - \frac{|\widehat{p}_{mnk} - \widehat{p}_{mn_{m}k}|}{\widehat{p}_{mn_{m}k}},
$$
where $m$ denotes the clicked bidder in round $n$.

\subsection{RL-based Payment Mechanisms}

Given the continuous action space, we employ an Actor-Critic algorithm based on proximal policy optimization (PPO) to learn per-click payment. The Actor-Critic framework comprises two components: the Actor, which selects actions, and the Critic, which evaluates the quality of those actions.

\textbf{Actor.} The objective of the Actor is to select actions that maximize the return $G_{n} = \sum_{j = 0}^{N - n}{\gamma^{j} r_{n + j}}$, where $\gamma$ denotes the discount factor governing the decay of future rewards. To facilitate effective policy updates, we introduce the advantage function $A_{n}^{(\gamma, \lambda)} = \sum_{j = 0}^{N - n}{(\gamma \lambda)^{j} \delta_{n + j}}$ as a metric that quantifies the deviation of a chosen action $\widehat{\boldsymbol{p}}_{n}$ from the average action in a given state $\boldsymbol{s}_{n}$. $A_{n}^{(\gamma, \lambda)}$ is calculated using generalized advantage estimation (GAE), $\gamma$ and $\lambda$ serve as discount factors, modulating the attenuation of rewards and advantages respectively. The temporal-difference error (TD), denoted by $\delta_{n} = r_{n} + \gamma V(s_{n+1}) - V(s_{n})$, captures the discrepancy between the immediate reward and the predicted change in state value, where $V(s_{n})$ is the output of the Critic, representing the state value function.

Within the Actor's iterative update process, the algorithm observes the behavior probability $\pi_{\theta} (\widehat{p}_{mnk}|s_{n})$ under the current policy at time $n$ and state $s_{n}$, and contrasts it with the corresponding probability $\pi_{\theta}^{old} (\widehat{p}_{mnk}|s_{n})$ under the previous policy. A probability ratio $\rho_{n}^{(\theta)} = \frac{\pi_{\theta} (\widehat{p}_{mnk}|s_{n})}{\pi_{\theta}^{old} (\widehat{p}_{mnk}|s_{n})}$ is subsequently computed to regulate the extent of policy update. This ratio plays a pivotal role in adjusting the update magnitude; if the difference between the new and old policies is significant and the advantage function is large, the update magnitude is appropriately increased. Conversely, a $\rho_{n}^{(\theta)}$ closer to 1 indicates a smaller difference between the policies. 

To maximize the advantage, we optimize the policy parameters $\theta$ in a manner that enhances the likelihood of actions associated with positive advantages while diminishing that of negative advantages. To achieve this, we employ the PPO-Clip method, which directly clips the ratio of the new and old policies, thereby constraining the magnitude of change. The loss function $L^{(\theta)}$ is:

$$
L^{(\theta)} = -\mathbf{E}_{n}\left[ \min \{ \rho_{n}^{(\theta)} A_{n}^{(\gamma, \lambda)}, \text{clip}(\rho_{n}^{(\theta)}, 1 - \kappa, 1 + \kappa) A_{n}^{(\gamma, \lambda)} \} \right],
$$
where $\kappa$ represents the clipping hyperparameter, typically set within the range $[0.1,0.2]$; $\text{clip}()$ ensures that $\rho_{n}^{(\theta)}$ remains within the interval $[1 - \kappa, 1 + \kappa]$ for convergence; and the $\min$ function selects the smaller value between the unclipped and clipped objectives, forming a lower bound on the target. The loss function $L^{(\theta)}$ considers two cases based on the sign of the advantage function $A_{n}^{(\gamma, \lambda)}$. If the advantage is positive, the policy ratio $\rho_{n}^{(\theta)}$ is increased, but no additional incentive is provided when $\rho_{n}^{(\theta)} > 1 + \kappa$. Conversely, if the advantage is negative, $\rho_{n}^{(\theta)}$ is decreased, but no additional penalty is imposed when $\rho_{n}^{(\theta)} < 1 - \kappa$, thus limiting the difference between the new and old policies to a reasonable range.

\textbf{Critic.} The Critic, serving as an evaluation network for the state value, takes the state as input, and provides the value estimates necessary for computing the advantages used by the Actor. The loss function $L^{\mu}$ of the Critic is computed using the mean squared error (MES) loss, which corresponds to the average of the cumulative squared TD error values. Specifically,

$$
L^{\mu} = \mathbf{E} \left[ \left( V^{(\mu)}(s_{n}) - G_{n} \right)^2 \right],
$$
where $\mu$ denotes the parameters of the Critic. The target value $G_{n}$ is calculated as the discounted sum of future rewards, given by $G_{n} = r_{n} + \gamma r_{n+1} + \cdots + \gamma^{N - n} r_{N} = \sum_{j=0}^{N-n}{\gamma^{j} r_{n+j}}$, where $\gamma$ is the discount factor.

During the training phase, combining the losses of the Actor and Critic, the overall loss function is formulated as

$$
L^{(\theta, \mu)} = \alpha_{1} L^{\theta} + \alpha_{2} L^{\mu} + \alpha_{3} H(\pi_{\theta}),
$$
where $\alpha_{1}, \alpha_{2}, \alpha_{3}$ are coefficients that control the weights of the respective terms. Specifically, $H(\pi_{\theta})$ represents the Shannon Entropy, and the term $\alpha_{3} H(\pi_{\theta})$ serves to balance the exploration and exploitation trade-off during the learning process.

\section{Experiment}
\label{experiment}

We conduct extensive experiments on real datasets to demonstrate our conclusions.

\subsection{Experimental Setup}
\label{setup}

To meticulously assess the properties and efficacy of \textbf{DFP}, this section presents some experimental framework leveraging genuine online data. Concretely, we have curated a dataset comprising 31 days of operational records from 5000 distinct bidders, encompassing hundreds of millions rounds of auctions. The dataset encompasses a rich array of metrics, including CTRs, CVRs, tCPA, click volume, and conversion volume for each bidder across every round. It is noteworthy that given the prior deployment of \textbf{DFP} within the live ad system (where the RL has been trained utilizing data from all bidders within the platform), we refrain from retraining the RL specifically for the experiments in this section. Instead, we directly use \textbf{DFP} that has been deployed in the system. This approach underscores the robustness and adaptability of our algorithm in dynamic and intricate real-world environments, as retraining for the test set would potentially mask the algorithm's ability to generalize. We empirically evaluate the performance of four mechanisms, including \textbf{CFP} and \textbf{DFP}, against two widely adopted auction mechanisms in autobidding as baselines. These mechanisms are introduced in Section \ref{intro}.

\begin{table*}[h]
\centering
\begin{tabular}{clll}
\hline
\multirow{2}{*}{Mechanism} & {\multirow{2}{*}{{\begin{tabular}[c]{@{}c@{}}\multicolumn{1}{c}{\textbf{$\frac{tCPA}{CPA}$}}\\  upper  \, lower  \, mean\end{tabular}}}} & {\multirow{2}{*}{{\begin{tabular}[c]{@{}c@{}}\multicolumn{1}{c}{$Var(p)$}\\  upper  \, lower  \, mean\end{tabular}}}} & {\multirow{2}{*}{{\begin{tabular}[c]{@{}c@{}}\multicolumn{1}{c}{$R(p)$}\\  upper  \, lower  \, mean\end{tabular}}}} \\
& \multicolumn{1}{c}{} & \multicolumn{1}{c}{} & \multicolumn{1}{c}{} \\ \hline
\textbf{CPA} & 1.000 \, 1.000 \, 1.000 & 0.073  \, 0.029  \, 0.056 & 1.000 \, 1.000 \, 1.000 \\ \hline
\textbf{Pacing} & 1.221 \, 0.787 \, 1.028 & 0.000  \, 0.000  \, 0.000 & 0.000 \, 0.000 \, 0.000 \\ \hline 
\textbf{CFP} & 1.176 \, 0.775 \,  \textbf{0.996} & 0.016  \, 0.008  \, 0.013 & 0.691 \, 0.468 \, 0.586 \\ \hline
\textbf{DFP} & \textbf{1.050} \, \textbf{0.917} \,  0.989 & \textbf{0.010}  \, \textbf{0.005}  \, \textbf{0.008} & \textbf{0.646} \, \textbf{0.421} \, \textbf{0.564} \\ \hline
\end{tabular}
\caption{Main experimental results.}\label{table1}
\end{table*}

\textbf{CPA}: We adopt the CPA pricing model within an offline setting, where the algorithm possesses complete knowledge of all auction outcomes, including clicks, conversions, CTRs, CVRs, and tCPA across all rounds and bidders. In optimization parlance, this offline setting represents the optimal solution to the online problem. Within this context, we utilize the allocation mechanism of \textbf{DFP} (which is identical to that of \textbf{CFP}) and apply the first-price payment mechanism in the CPA model. Precisely, if a click fails to result in a conversion, the platform charges 0; otherwise, it bills the bidder for her bid (i.e., $tCPA$), denoted as $\widehat{p}_{mnk} = \widehat{z}_{mnk} \cdot tCPA_{m}$. For brevity, we refer to this baseline as \textbf{CPA}.

\textbf{Pacing}: Inspired by Aggarwal et al.\cite{aggarwal2019autobidding}, we configure the autobidders to participate in auctions using a pacing bidding strategy (akin to uniform bidding). We also evaluate this baseline within the offline setting, where bidders leverage complete online information to compute the optimal uniform bid. Essentially, the optimal uniform bid evenly distributes the total payments across clicks, resulting in a pacing factor of 1. Thus, the payment is $\widehat{p}_{mnk} = \widehat{y}_{mnk} \cdot \frac{\widehat{Z}_{mN} \cdot tCPA_{m}}{\widehat{Y}_{mN}}$. For simplicity, we designate this baseline as \textbf{Pacing}.

To evaluate the performance of these mechanisms, we rely on two pivotal metrics: TIC and payment fluctuation. Drawing upon Lemma \ref{lemma4}, we quantify TIC using the ratio $\frac{tCPA}{CPA}$.

\subsection{Performance Evaluation}
\label{evaluation}

\subsubsection{TIC}
\label{exptic}

\begin{figure}[h]
\centering
\includegraphics[width=0.5\textwidth]{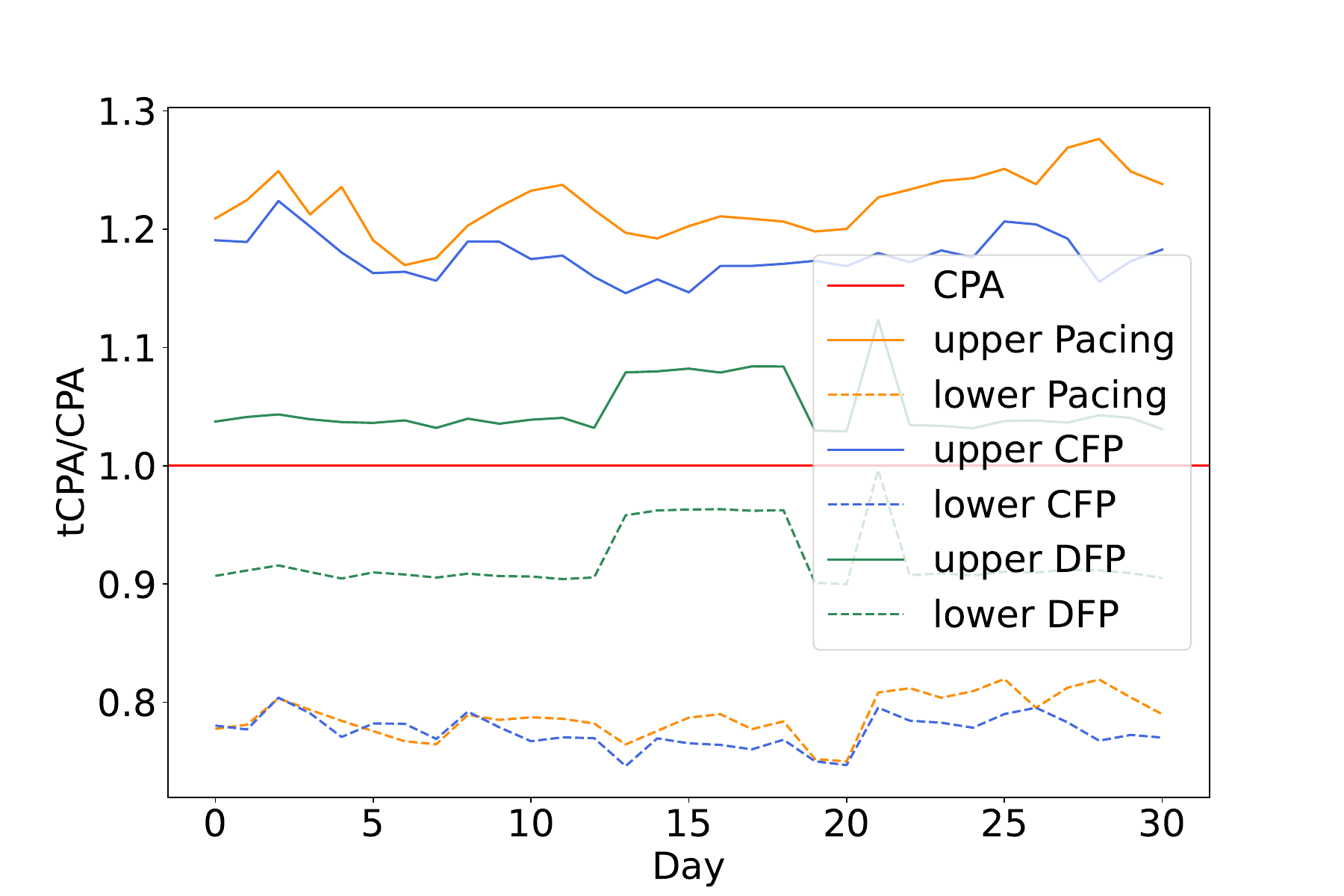}
\caption{The $\frac{tCPA}{CPA}$ ratio of 5000 bidders over 31 days.}
\label{result1}
\end{figure}

We partition the 31-day period into 31 distinct stages, each stage corresponding to a single day. Utilizing the dataset in Section \ref{setup}, we ran four distinct mechanisms and obtained $CPA_{mN_t}$ for each bidder $m$ in stage (day) $t$. For each mechanism, we computed $31 \times 5000 = 155,000$ instances of $\frac{tCPA_m}{CPA_{mN_t}}$. To mitigate the influence of noisy points, we calculated the upper and lower quartiles, as well as the mean, of these $155,000$ ratios for each mechanism, and recorded the results in Table \ref{table1}. Furthermore, at each stage, we identified the upper and lower quartiles among these $5000$ ratios, and plotted these findings in Figure \ref{result1}.

The experimental results demonstrate that \textbf{CPA} is TIC, since \textbf{CPA} operates under an offline setting, representing the optimal solution among all online mechanisms. The remaining three mechanisms are $\epsilon$-TIC, with \textbf{DFP} exhibiting the closest ratio of $\frac{tCPA}{CPA}$ to 1, thus incurring the smallest $\epsilon_{DFP}$. This empirical evidence underscores that in the online setting, \textbf{DFP} is the mechanism that most closely approximates TIC.

\subsubsection{AIC and TIC}

\begin{figure}[h]
\centering
\includegraphics[width=0.5\textwidth]{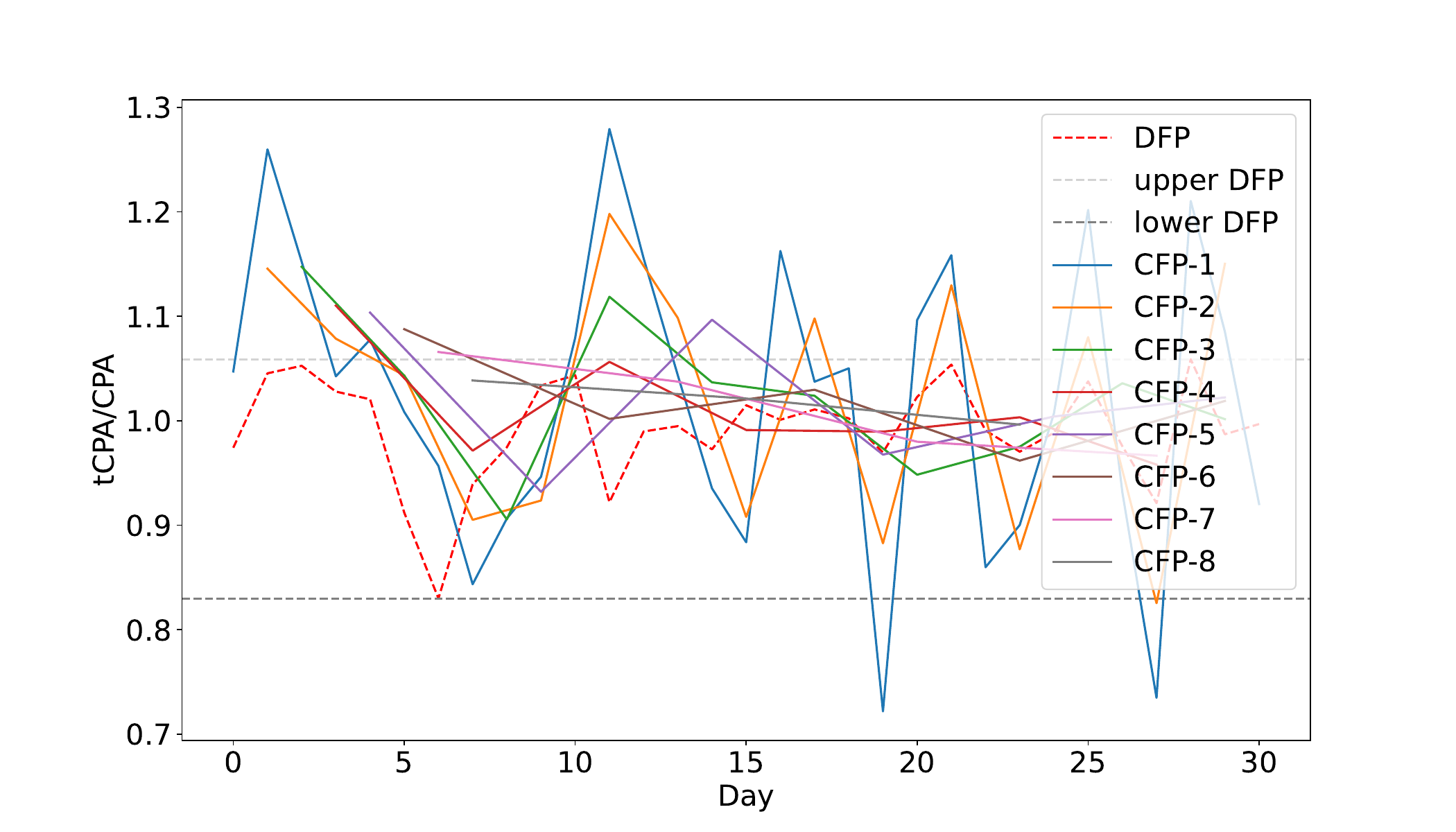}
\caption{Comparison between \textbf{CFP} and \textbf{DFP}.}
\label{result2}
\end{figure}

To further compare \textbf{CFP} and \textbf{DFP}, we denote $\textbf{CFP-}\tau$ as the \textbf{CFP} with $\tau$ days per stage. We believe that as $\tau$ increases, the aggregate click volume per stage approaches the threshold in Lemma \ref{lemma5}, thereby bringing the performance of \textbf{CFP} closer to that of \textbf{DFP}. To validate this hypothesis, we randomly sample an bidder from the dataset in Section \ref{setup} and run $\textbf{CFP-}\tau$. The result are presented in Figure \ref{result2}.

Our experiments reveal that \textbf{CFP-8}, with eight days per stage, achieves performance that is comparable to \textbf{DFP}. Notably, this bidder gets an average of 1626 clicks every 8 days, a substantial figure that few bidders in the dataset of Section \ref{setup} can achieve such average daily click volume. This experiment highlights the need for a decoupled auction mechanism.

\subsubsection{Sparse-Click Systems}
\label{expsps}

\begin{figure}[h]
\centering
\includegraphics[width=0.5\textwidth]{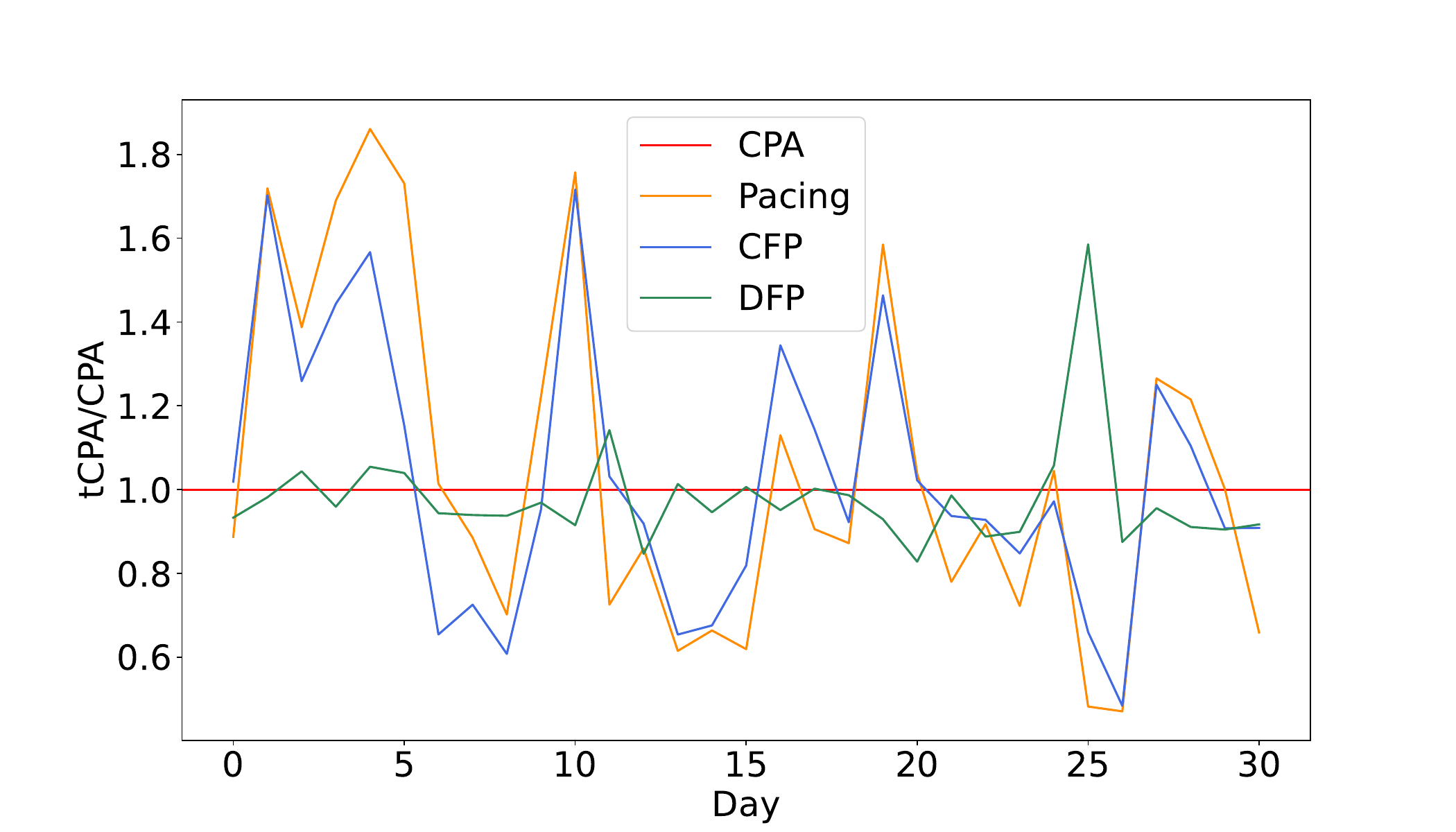}
\caption{The $\frac{tCPA}{CPA}$ ratio of a bidder with an average daily click volume of 132.}
\label{result4}
\end{figure}

To validate the superiority of \textbf{DFP} in sparse-click systems, we initiated by identifying a set of bidders whose daily click volume fell within the range of 100 to 150, from which a single bidder was randomly sampled. Subsequently, we implemented four distinct mechanisms and plotted the sampled bidder's $\frac{tCPA}{CPA}$ ratio in Figure \ref{result4}.

Notably, even though the bidder's average daily click volume is significantly below the threshold stipulated in Lemma \ref{lemma5}, the condition $\epsilon_{DFP} \le 0.1$ held almost invariably, with $\epsilon_{DFP}$ notably smaller than $\epsilon_{CFP}$ and $\epsilon_{Pacing}$. This underscores the fact that \textbf{DFP} remains nearly TIC even in highly sparse-click systems.

\subsubsection{Payment Fluctuation}
\label{exppay}

To enhance the payment experience for bidders and mitigate excessive variations in per-click payments, we incorporated a penalty term (negative reward) within the reward function of our RL algorithm. This penalty was designed to suppress fluctuations in per-click payments. Utilizing the dataset in Section \ref{setup}, we ran four mechanisms and recorded the post-click payment $\widehat{p}_{mnk}$ for each bidder $m$. For each bidder $m$, we filtered out impressions that did not result in clicks and subsequently normalized all payments $\widehat{p}_{mnk}$ to the interval $[0, 1]$: $\widehat{p}_{mnk} \leftarrow \frac{\widehat{p}_{mnk}}{tCPA_m}$.

This normalization step was applied across all bidders within each mechanism, enabling a fair comparison of outcomes across mechanisms on a consistent scale. Post-normalization, we computed the variance ($Var$) and range ($R$) of all post-click payments for each bidder $m$, resulting in 5000 variance values and 5000 range values. For each mechanism, we documented the upper and lower quartiles as well as the mean of these variance and range metrics, with the results summarized in Table \ref{table1}.

Our experimental results indicate that in \textbf{Pacing}, the per-click payments of bidders exhibit no fluctuation, which is intuitive given that \textbf{Pacing} operates in an offline setting, enabling the utilization of an optimal uniform bidding strategy for both bidding and payment. Conversely, \textbf{CPA} demonstrates the most significant fluctuation, attributed to its binary payment nature where payments are either 0 or tCPA. The fluctuation levels of \textbf{CFP} and \textbf{DFP} lie between these extremes, with \textbf{DFP} exhibiting lower fluctuation than \textbf{CFP}. These findings align with our expectations, corroborating that \textbf{DFP} effectively dampens payment fluctuations and, in doing so, mitigates the uncertainty associated with $\widehat{Z}_{mN_t}$.

The aforementioned experiments validate the primary conclusion of this paper: \textbf{DFP} is TIC. Furthermore, we conducted additional experiments to demonstrate the advantages of \textbf{CFP} and \textbf{DFP} in other aspects. Due to space constraints, these supplementary experiments are detailed in Appendix \ref{addexps}.

\section{Conclusion and Future Work}
\label{conclusion}

In this paper, we devise an AIC mechanism, \textbf{CFP}. To address the phenomenon of untruthful bidding by risk-averse bidders in AIC mechanisms, we decouple the allocation and payment mechanism of \textbf{CFP} and design a TIC mechanism, \textbf{DFP}, in which risk-averse bidders bid truthfully. Then, using RL, we propose a complete closed-loop process from theoretical foundation to practical implementation of \textbf{DFP}. In future work, we aim to leverage the decoupling technique we proposed to devise TIC mechanisms for other pricing models within autobidding frameworks and accommodate various constraints of advertisers.


\bibliographystyle{unsrt}  
\bibliography{main}  

\newpage

\appendix

\section{Additional Experiments}
\label{addexps}

\subsection{Sparse-Click Systems}

\begin{figure}[h]
\centering
\includegraphics[width=0.45\textwidth]{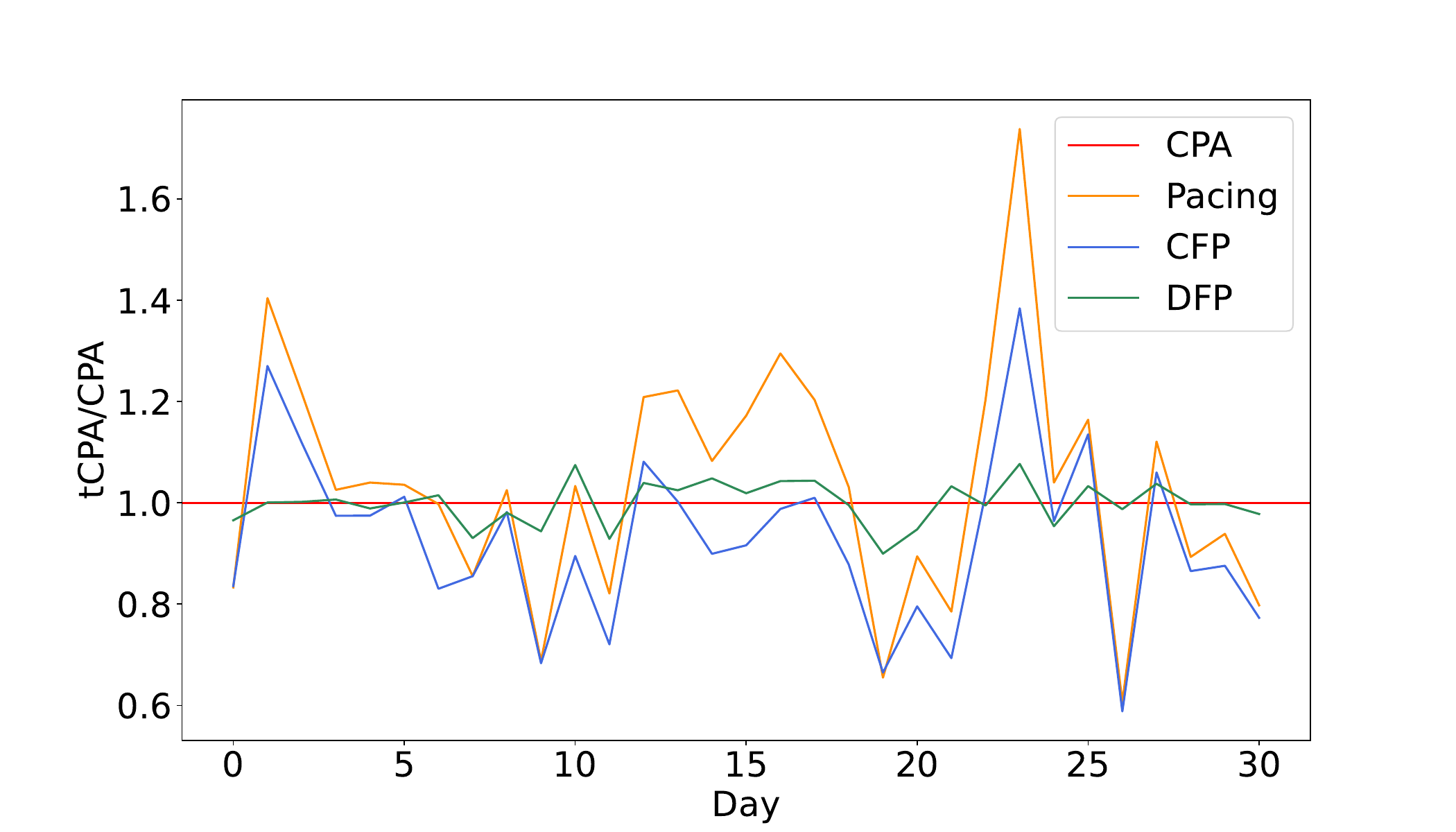}
\caption{The $\frac{tCPA}{CPA}$ ratio of a bidder with an average daily click volume of 1308.}
\label{result3}
\end{figure}

To validate the superiority of \textbf{DFP} in sparse-click systems, we initiated by identifying a set of bidders whose daily click volume fell within the range of 1000 to 1500, from which a single bidder was randomly sampled. Subsequently, we implemented four distinct mechanisms and plotted the sampled bidder's $\frac{tCPA}{CPA}$ ratio in Figure \ref{result3}.

Notably, even though the bidder's average daily click volume is significantly below the threshold stipulated in Lemma \ref{lemma5}, the condition $\epsilon_{DFP} \le 0.1$ held almost invariably, with $\epsilon_{DFP}$ notably smaller than $\epsilon_{CFP}$ and $\epsilon_{Pacing}$. This underscores the fact that \textbf{DFP} remains nearly TIC even in highly sparse-click systems. Furthermore, we observe that in Figure \ref{result4}, the performance of \textbf{CFP} and \textbf{Pacing} is comparable, whereas in Figure \ref{result3}, while \textbf{CFP} underperforms compared to \textbf{DFP}, it surpasses \textbf{Pacing}. Based on these findings, we can prove the following conclusion: As the click volume transitions from sparse to dense, the performance of \textbf{CFP} gradually converges towards that of the TIC mechanism.

\subsection{Payment Fluctuation}

\begin{figure}[h]
\centering
\includegraphics[width=0.5\textwidth]{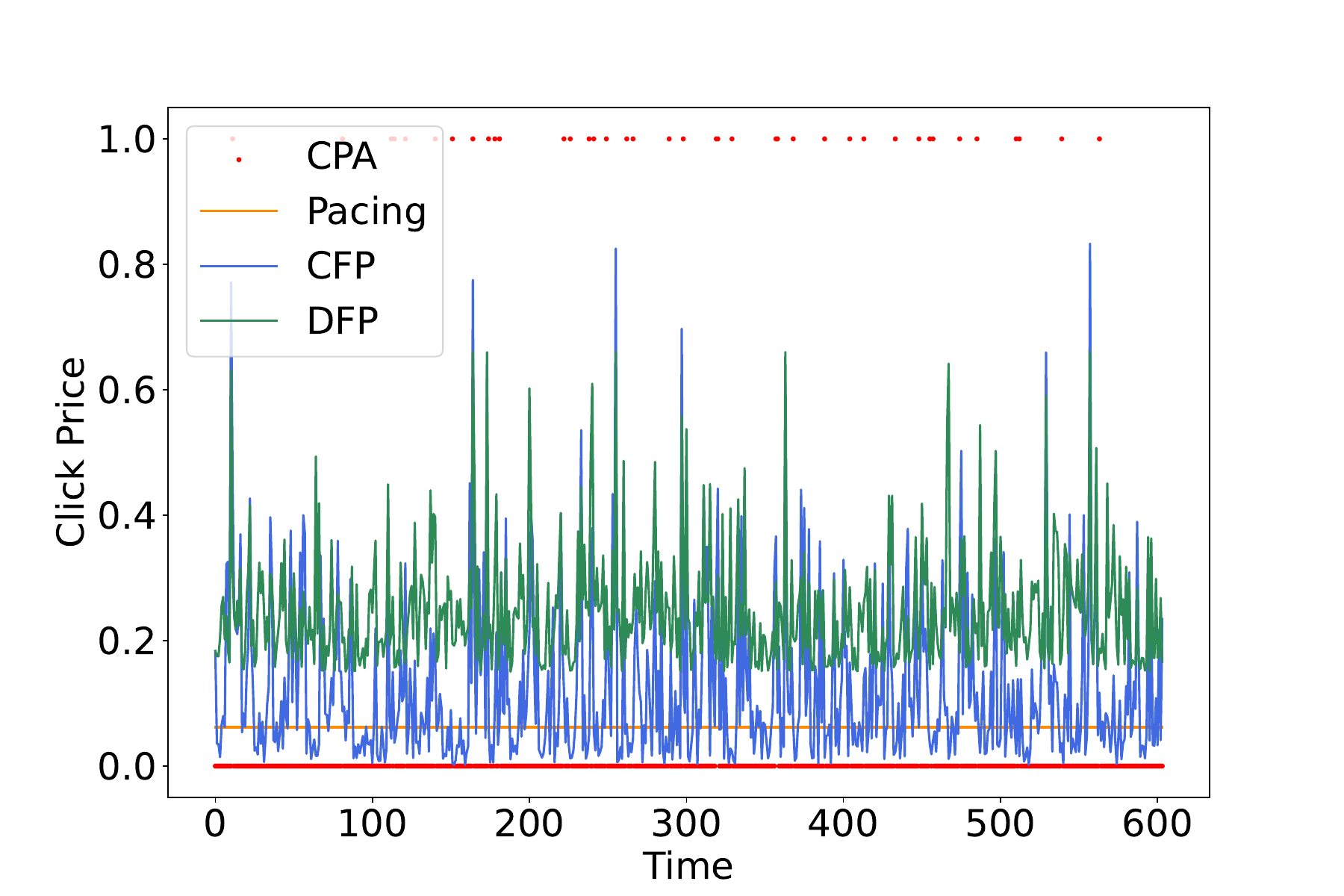}
\caption{Per-click payments of the sampled bidder over 31 days.}
\label{result5}
\end{figure}

Utilizing the dataset described in Section \ref{setup}, we randomly sampled an bidder and subsequently executed four distinct mechanisms, recording the per-click payments of this bidder under each mechanism. The results of these experiments are depicted in Figure \ref{result5}.

Our experimental findings reveal that in the \textbf{Pacing} mechanism, the bidder's per-click payments exhibit no fluctuations, which is intuitive given that \textbf{Pacing} operates in an offline setting, allowing for the application of an optimal uniform bidding strategy for both bidding and payment. In contrast, the \textbf{CPA} mechanism demonstrates the most significant fluctuation, attributed to its binary payment scheme where payments are either 0 or tCPA. The fluctuation levels of \textbf{CFP} and \textbf{DFP} lie between these extremes, with \textbf{DFP} exhibiting a lower degree of fluctuation compared to \textbf{CFP}. These results align with our expectations, confirming that \textbf{DFP} is effective in suppressing payment fluctuations, thereby mitigating the uncertainty associated with $\widehat{Z}_{mN_t}$.

\section{Chernoff Bound}
\label{chernoff}

\textbf{Sums of independent Bernoulli random variables}

The bounds in the following sections for Bernoulli random variables are derived by using that, for a Bernoulli random variable $X_{i}$ with probability $p$ of being equal to $1$,

$$
\mathbf{E}\left[e^{t\cdot X_{i}}\right] = (1-p) e^{0} + p e^{t} = 1 + p(e^{t} - 1) \le e^{p(e^{t} - 1)}.
$$

One can encounter many flavors of Chernoff bounds: the original additive form (which gives a bound on the absolute error) or the more practical multiplicative form (which bounds the error relative to the mean).

\textbf{Multiplicative form (relative error)}

Multiplicative Chernoff bound. Suppose $X_{1}, \dots, X_{n}$ are independent random variables taking values in $\{0, 1 \}$. Let $X$ denote their sum and let $\mu = E[X]$ denote the sum's expected value. Then for any $\epsilon > 0$,

$$
\mathbf{Pr}(X \ge (1 + \epsilon) \mu) \le \left( \frac{e^{\epsilon}}{(1 + \epsilon)^{1 + \epsilon}} \right)^{\mu}. 
$$

A similar proof strategy can be used to show that for $0 < \epsilon < 1$,

$$
\mathbf{Pr}(X \le (1 - \epsilon) \mu) \le \left( \frac{e^{-\epsilon}}{(1 - \epsilon)^{1 - \epsilon}} \right)^{\mu}. 
$$

The above formula is often unwieldy in practice, so the following looser but more convenient bounds are often used, which follow from the inequality $\frac{2 \epsilon}{2 + \epsilon} \le \log{(1 + \epsilon)}$ from the list of logarithmic inequalities:

\begin{align*}
& \mathbf{Pr}(X \ge (1 + \epsilon) \mu) \le e^{-\epsilon^{2} \mu / (2 + \epsilon)}, \quad 0 \le \epsilon, \\
& \mathbf{Pr}(X \le (1 - \epsilon) \mu) \le e^{-\epsilon^{2} \mu / 2}, \quad 0 \le \epsilon, \\
& \mathbf{Pr}(|X - \mu| \ge \epsilon \mu) \le 2e^{-\epsilon^{2} \mu / 3}, \quad 0 \le \epsilon.
\end{align*}

Notice that the bounds are trivial for $\epsilon = 0$.

\section{Missing Proofs}
\label{proof}

\subsection{Proof of Lemma \ref{lemma1}}

\begin{proof}

By the definition of $\overline{u}_{mN}(b_{m}, \boldsymbol{b}_{-m})$, we have

\begin{align*}
& \overline{u}_{mN}(b_{m}, \boldsymbol{b}_{-m}) = \sum_{n \in [N], k \in [K]}{v_{mnk} \cdot \overline{z}_{mnk}} \\
= & \sum_{n \in [N], k \in [K]}{v_{mnk} \cdot \overline{y}_{mnk} \cdot cvr_{mnk}} \\
= & \sum_{n \in [N], k \in [K]}{v_{mnk} \cdot x_{mnk}(b_{m}, \boldsymbol{b}_{-m}) \cdot ctr_{mnk} \cdot cvr_{mnk}}.
\end{align*}

Based on the definition of the \textbf{CFP} mechanism, it is established that $\frac{\partial x_{mnk}(b_{m}, \boldsymbol{b}_{-m})}{\partial b_{m}} \ge 0$. As the sum of several weakly increasing functions remains weakly increasing, it follows that $\frac{\partial \overline{u}_{mN}(b_{m}, \boldsymbol{b}_{-m})}{\partial b_{m}} \ge 0$. Consequently, under the tCPA constraint (i.e., $CPA_{mN} \leq tCPA_{m}$), bidder $m$ should aim to maximize their bid $b_{m}$ to the extent possible. By the definition of $\overline{CPA}_{mN}$,

\begin{align*}
& \overline{CPA}_{mN} = \frac{\overline{P}_{mN}}{\overline{Z}_{mN}} = \frac{\sum_{n \in [N], k \in [K]}{\overline{p}_{mnk}}}{\sum_{n \in [N], k \in [K]}{\overline{z}_{mnk}}} \\
= & \frac{\sum_{n \in [N], k \in [K]}{b_{m} \cdot \overline{y}_{mnk} \cdot cvr_{mnk}}}{\sum_{n \in [N], k \in [K]}{\overline{y}_{mnk} \cdot cvr_{mnk}}} \\
= & \frac{\sum_{n \in [N], k \in [K]}{b_{m} \cdot x_{mnk} \cdot ctr_{mnk} \cdot cvr_{mnk}}}{\sum_{n \in [N], k \in [K]}{x_{mnk} \cdot ctr_{mnk} \cdot cvr_{mnk}}} = b_{m}.
\end{align*}

Ad $m$ should maximize the bid $b_{m}$ subject to the constraint $b_{m} \leq tCPA_{m}$. Therefore, $b_{m}^{*} = tCPA_{m}$ is the optimal bidding strategy for the value-maximizer $m$ with a tCPA constraint.

When $b_{m} = tCPA_{m}$,

\begin{align*}
& \overline{CPA}_{mN} = \frac{\overline{P}_{mN}}{\overline{Z}_{mN}} = \frac{\sum_{n \in [N], k \in [K]}{\overline{p}_{mnk}}}{\sum_{n \in [N], k \in [K]}{\overline{z}_{mnk}}} \\
= & \frac{\sum_{n \in [N], k \in [K]}{tCPA_{m} \cdot \overline{y}_{mnk} \cdot cvr_{mnk}}}{\sum_{n \in [N], k \in [K]}{\overline{y}_{mnk} \cdot cvr_{mnk}}} \\
= & \frac{\sum_{n \in [N], k \in [K]}{tCPA_{m} \cdot x_{mnk} \cdot ctr_{mnk} \cdot cvr_{mnk}}}{\sum_{n \in [N], k \in [K]}{x_{mnk} \cdot ctr_{mnk} \cdot cvr_{mnk}}} \\
= & tCPA_{m} \le tCPA_{m}.
\end{align*}

Therefore, $b_{m} = tCPA_{m}$ does not violate the tCPA constraint of bidder $m$.

Extending this argument, if all bidders are value-maximizers with tCPA constraints, each bidder $m$ faces an optimal bidding problem $\overline{\boldsymbol{P}}_{m}$. Consequently, $b_{m}^{*} = tCPA_{m}, \forall m \in [M]$ holds. Therefore, if all bidders are value-maximizers with tCPA constraints, $b_{m}^{*} = tCPA_{m}, \forall m \in [M]$ constitutes a Nash Equilibrium.
    
\end{proof}

\subsection{Proof of Theorem \ref{theorem1}}

\begin{proof}

In the \textbf{CFP} mechanism, if all bidders are value-maximizers with tCPA constraints, the strategy $b_{m}^{*} = tCPA_{m}, \forall m \in [M]$ constitutes a Nash Equilibrium. This implies that truthfully reporting one's tCPA as the bidding price is a dominant strategy for all bidders, and truthful bidding will not violate her tCPA constraint. Therefore, the \textbf{CFP} mechanism satisfies the properties of AIC and IR.
    
\end{proof}

\subsection{Proof of Lemma \ref{lemma2}}

\begin{proof}

$\mathcal{M}(\boldsymbol{x}, \boldsymbol{p})$ is AIC and IR $\Longleftrightarrow$ $\overline{CPA}_{mN} = tCPA_{m}$.

\begin{enumerate}

\item[$\Longrightarrow$]:

If $\mathcal{M}(\boldsymbol{x}, \boldsymbol{p})$ is AIC and IR, then $b_{m}^{*} = tCPA_{m}$. Substitute $b_{m} = tCPA_{m}$ into $\overline{CPA}_{mN}$,

\begin{align*}
& \overline{CPA}_{mN} = \frac{\overline{P}_{mN}}{\overline{Z}_{mN}} = \frac{\sum_{n \in [N], k \in [K]}{\overline{p}_{mnk}}}{\sum_{n \in [N], k \in [K]}{\overline{z}_{mnk}}} \\
= & \frac{\sum_{n \in [N], k \in [K]}{tCPA_{m} \cdot \overline{y}_{mnk} \cdot cvr_{mnk}}}{\sum_{n \in [N], k \in [K]}{\overline{y}_{mnk} \cdot cvr_{mnk}}} \\
= & \frac{\sum_{n \in [N], k \in [K]}{tCPA_{m} \cdot x_{mnk} \cdot ctr_{mnk} \cdot cvr_{mnk}}}{\sum_{n \in [N], k \in [K]}{x_{mnk} \cdot ctr_{mnk} \cdot cvr_{mnk}}} = tCPA_{m}.
\end{align*}

\item[$\Longleftarrow$]:

When $\overline{CPA}_{mN} = tCPA_{m}$, according to the definition of $\overline{CPA}_{mN}$, we have

\begin{align*}
& \overline{CPA}_{mN} = \frac{\overline{P}_{mN}}{\overline{Z}_{mN}} = \frac{\sum_{n \in [N], k \in [K]}{\overline{p}_{mnk}}}{\sum_{n \in [N], k \in [K]}{\overline{z}_{mnk}}} \\
= & \frac{\sum_{n \in [N], k \in [K]}{b_{m} \cdot \overline{y}_{mnk} \cdot cvr_{mnk}}}{\sum_{n \in [N], k \in [K]}{\overline{y}_{mnk} \cdot cvr_{mnk}}} \\
= & \frac{\sum_{n \in [N], k \in [K]}{b_{m} \cdot x_{mnk} \cdot ctr_{mnk} \cdot cvr_{mnk}}}{\sum_{n \in [N], k \in [K]}{x_{mnk} \cdot ctr_{mnk} \cdot cvr_{mnk}}} = b_{m}.
\end{align*}

Therefore, $b_{m} = tCPA_{m}$, which means that bidders will bid truthfully, that is, $\mathcal{M}(\boldsymbol{x}, \boldsymbol{p})$ is AIC and IR.

\end{enumerate}
    
\end{proof}

\subsection{Proof of Lemma \ref{lemma3}}

\begin{proof}

\begin{enumerate}

\item[$\Longrightarrow$]:

If truthful bidding constitutes the optimal strategy for advertisement $m$, then the mechanism $\mathcal{M}$ is AIC. According to Lemma \ref{lemma2}, which states that $\mathcal{M}$ is AIC if and only if $\overline{CPA}_{mn} = tCPA_m$. Consequently, a risk-averse bidder will truthfully bid iff $\overline{CPA}_{mn} = tCPA_m$. By definition, the risk-averse bidder believes that the actual outcome should be equal to the expectation, that is, $\widehat{CPA}_{mN_t} = \overline{CPA}_{mN_t}, \forall t \in [T]$. Hence, if the risk-averse bidder bids truthfully, it follows that $\frac{tCPA_m}{\widehat{CPA}_{mN_t}} = 1, \forall t \in [T]$.

\item[$\Longleftarrow$]:

If $\frac{tCPA_m}{\widehat{CPA}_{mN_t}} = 1, \forall t \in [T]$, by definition, a risk-averse bidder believes that the actual outcome should be equal to the expectation, that is, $\widehat{CPA}_{mN_t} = \overline{CPA}_{mN_t}, \forall t \in [T]$. Subsequently, it can be inferred that $\overline{CPA}_{mN_t} = tCPA_m, \forall t \in [T]$. Leveraging Lemma \ref{lemma2} again, we confirm that the mechanism $\mathcal{M}$ is AIC if and only if $\overline{CPA}_{mn} = tCPA_m$, and truthful bidding is indeed the optimal strategy for advertisement $m$ within an AIC mechanism. Thus, when $\frac{tCPA_m}{\widehat{CPA}_{mN_t}} = 1, \forall t \in [T]$, it confirms that a risk-averse bidder will bid truthfully.

\end{enumerate}

\end{proof}

\subsection{Proof of Lemma \ref{lemma4}}

\begin{proof}

According to Lemma \ref{lemma3}, in mechanism $\mathcal{M}$, a risk-averse bidder will bid truthfully during the auction process iff $\frac{tCPA_m}{\widehat{CPA}_{mN_t}} = 1, \forall t \in [T]$.

Furthermore, as stated in Definition \ref{definition6}, a mechanism $\mathcal{M}$ is Time-Invariant Incentive Compatibility iff truthful bidding is the optimal strategy for all risk-averse bidders during the auction process.

Consequently, we conclude that, in OCPC model with tCPA ads, a mechanism $\mathcal{M}$ is TIC iff $\frac{tCPA_m}{\widehat{CPA}_{mN_t}} = 1, \forall m \in [M],  \forall t \in [T]$.
    
\end{proof}

\subsection{Proof of Lemma \ref{lemma5}}

\begin{proof}

By definition, for the mechanism $\mathcal{M}(\boldsymbol{x}, \boldsymbol{p})$ to satisfy the $\epsilon$-TIC property, it is required that

$$
1 - \epsilon \le \frac{tCPA_m}{\widehat{CPA}_{mN_t}} \le 1 + \epsilon, \forall t \in [T].
$$

By the definition of $\widehat{CPA}_{mN_t}$, this is equivalent to

$$
\epsilon\text{-TIC} \Leftrightarrow 1 - \epsilon \le \frac{\widehat{Z}_{mn} \cdot tCPA_m}{\widehat{P}_{mn}} \le 1 + \epsilon, \forall t \in [T].
$$

In the mechanism $\mathcal{M}(\boldsymbol{x}, \boldsymbol{p})$, the actual payment $\widehat{P}_{mn}$ equals the expected payment $\overline{P}_{mn}$, but the actual conversions $\widehat{Z}_{mn}$ may not necessarily equal the expected conversions $\overline{Z}_{mn}$. Therefore,

\begin{align*}
& \epsilon\text{-TIC} \\
\Leftrightarrow & 1 - \epsilon \le \frac{\widehat{Z}_{mn} \cdot tCPA_m}{\sum_{n \in [N_{t}], k \in [K]}{tCPA_m \cdot \widehat{y}_{mnk} \cdot cvr_{mnk} }} \le 1 + \epsilon \\
\Leftrightarrow & 1 - \epsilon \le \frac{\widehat{Z}_{mn}}{\sum_{n \in [N_{t}], k \in [K]}{\widehat{y}_{mnk} \cdot cvr_{mnk}}} \le 1 + \epsilon, \forall t \in [T].
\end{align*}

Next, we analyze the conditions on $N_{t}$ that ensure the above event holds.

\begin{align*}
& \mathbf{Pr}\left[1 - \epsilon \le \frac{\widehat{Z}_{mn}}{\sum_{n \in [N_{t}], k \in [K]}{\widehat{y}_{mnk} \cdot cvr_{mnk}}}\right] \\
= & \mathbf{Pr}\left[\widehat{Z}_{mn} \ge (1 - \epsilon) \sum_{n \in [N_{t}], k \in [K]}{\widehat{y}_{mnk} \cdot cvr_{mnk}}\right] \\
= & 1 - \mathbf{Pr}\left[\widehat{Z}_{mn} \le (1 - \epsilon) \sum_{n \in [N_{t}], k \in [K]}{\widehat{y}_{mnk} \cdot cvr_{mnk}}\right].
\end{align*}

The last equality is a direct consequence of the complementarity of probabilities. Using the Chernoff Bound (see Appendix \ref{chernoff}), we have

\begin{align*}
& \mathbf{Pr}\left[\widehat{Z}_{mn} \leq (1 - \epsilon) \sum_{n \in [N_{t}], k \in [K]}{\widehat{y}_{mnk} \cdot cvr_{mnk}}\right] \\
\le & \exp\left\{ - \frac{\epsilon^{2} \sum_{n \in [N_{t}], k \in [K]}{\widehat{y}_{mnk} \cdot cvr_{mnk}}}{2} \right\}.
\end{align*}

To ensure that the mechanism $\mathcal{M}(\boldsymbol{x}, \boldsymbol{p})$ satisfies $\epsilon$-TIC w.p. of at least $1 - \epsilon$, it is required that

$$
\exp\left\{ - \frac{\epsilon^{2} \sum_{n \in [N_{t}], k \in [K]}{\widehat{y}_{mnk} \cdot cvr_{mnk}}}{2} \right\} \leq \epsilon.
$$

which implies

$$
\sum_{n \in [N_{t}], k \in [K]}{\widehat{y}_{mnk} \cdot cvr_{mnk}} \geq \frac{2 \ln\left(\frac{1}{\epsilon}\right)}{\epsilon^2}.
$$

Let $\eta = \max_{m \in [M], n \in [N], k \in [K]} \{ cvr_{mnk} \}$, we can derive $\widehat{Y}_{mN_{t}} \ge \frac{2 \ln\left(\frac{1}{\epsilon}\right)}{\epsilon^2 \cdot \eta}$.

Similarly,

\begin{align*}
& \mathbf{Pr}\left[\frac{\widehat{Z}_{mn}}{\sum_{n \in [N_{t}], k \in [K]}{\widehat{y}_{mnk} \cdot cvr_{mnk}}} \leq 1 + \epsilon\right] \\
= & \mathbf{Pr}\left[\widehat{Z}_{mn} \le (1 + \epsilon) \sum_{n \in [N_{t}], k \in [K]}{\widehat{y}_{mnk} \cdot cvr_{mnk}}\right] \\
= & 1 - \mathbf{Pr}\left[\widehat{Z}_{mn} \ge (1 + \epsilon) \sum_{n \in [N_{t}], k \in [K]}{\widehat{y}_{mnk} \cdot cvr_{mnk}}\right].
\end{align*}

The last equality is a direct consequence of the complementarity of probabilities. Using the Chernoff Bound (see Appendix \ref{chernoff}), we have

\begin{align*}
& \mathbf{Pr}\left[\widehat{Z}_{mn} \ge (1 + \epsilon) \sum_{n \in [N_{t}], k \in [K]}{\widehat{y}_{mnk} \cdot cvr_{mnk}}\right] \\
\le & \exp\left\{ - \frac{\epsilon^{2} \sum_{n \in [N_{t}], k \in [K]}{\widehat{y}_{mnk} \cdot cvr_{mnk}}}{2 + \epsilon} \right\}.
\end{align*}

To ensure that the mechanism $\mathcal{M}(\boldsymbol{x}, \boldsymbol{p})$ satisfies $\epsilon$-TIC w.p. of at least $1 - \epsilon$, it is required that

$$
\exp\left\{ - \frac{\epsilon^{2} \sum_{n \in [N_{t}], k \in [K]}{\widehat{y}_{mnk} \cdot cvr_{mnk}}}{2 + \epsilon} \right\} \leq \epsilon.
$$

which implies

$$
\sum_{n \in [N_{t}], k \in [K]}{\widehat{y}_{mnk} \cdot cvr_{mnk}} \geq \frac{(2 + \epsilon) \ln\left(\frac{1}{\epsilon}\right)}{\epsilon^2}.
$$

From this, we can derive $\widehat{Y}_{mN_{t}} \geq \frac{(2 + \epsilon) \ln\left(\frac{1}{\epsilon}\right)}{\epsilon^2 \cdot \eta}$.

\end{proof}

\subsection{Proof of Lemma \ref{lemma6}}

\begin{proof}

According to Lemma \ref{lemma4}, the \textbf{CFP} mechanism is TIC iff $\frac{tCPA_m}{\widehat{CPA}_{mN_t}} = 1, \forall m \in [M], \forall t \in [T]$. By definition,

$$
\frac{tCPA_m}{\widehat{CPA}_{mN_t}} = \frac{\widehat{Z}_{mN_t} \cdot tCPA_m}{\widehat{P}_{mN_t}}= \frac{\widehat{Z}_{mN_t} \cdot tCPA_m}{\sum_{n \in [N_t], k \in [K]}{\widehat{y}_{mnk} \cdot \widehat{p}_{mnk}}}.
$$

Therefore, the \textbf{CFP} mechanism is TIC iff $\widehat{Z}_{mN_t} \cdot tCPA_{m} = \sum_{n \in [N_t], k \in [K]}{\widehat{y}_{mnk} \cdot \widehat{p}_{mnk}}, \forall m, t$.

\end{proof}

\subsection{Proof of Theorem \ref{theorem2}}

\begin{proof}

According to the definition of \textbf{DFP}, substituting $\widehat{p}_{mnk} = p_{mnk}^{*}$ into $\widehat{\boldsymbol{P}}_{Plt}$, we have

$$
\sum_{m \in [M]}{\left| \frac{\widehat{Z}_{mN_t} \cdot tCPA_{m}}{\sum_{n \in [N_t], k \in [K]}{\widehat{y}_{mnk} \cdot \widehat{p}_{mnk}}} - 1 \right|} = 0.
$$

Using the conclusion of Lemma \ref{lemma6}, we can get that \textbf{DFP} is TIC in OCPC model with tCPA ads.
    
\end{proof}

\end{document}